# Title: Mouse V1 population correlates of visual detection rely on heterogeneity within neuronal response patterns

**Authors**

Authors: Jorrit S. Montijn[1], Pieter M. Goltstein[1,2], Cyriel M.A. Pennartz[1,3]

1 Swammerdam Institute for Life Sciences, Center for Neuroscience, Faculty of Science, University of Amsterdam, the Netherlands

2 Max Planck Institute of Neurobiology, Martinsried, Germany

3 Research Priority Program Brain and Cognition, University of Amsterdam, the Netherlands

**Contact Information**

Correspondence and requests for materials should be addressed to C.M.A.P. (c.m.a.pennartz@uva.nl) or to J.S.M. (j.s.montijn@uva.nl).

**Abstract**

Previous studies have demonstrated the importance of the primary sensory cortex for the detection, discrimination and awareness of visual stimuli, but it is unknown how neuronal populations in this area process detected and undetected stimuli differently. Critical differences may reside in the mean strength of responses to visual stimuli, as reflected in bulk signals detectable in fMRI, EEG or MEG studies, or may be more subtly composed of differentiated activity of individual sensory neurons. Quantifying single-cell $Ca^{2+}$ responses to visual stimuli recorded with *in vivo* 2-photon imaging, we found that visual detection correlates more strongly with population response heterogeneity rather than overall response strength. Moreover, neuronal populations showed consistencies in activation patterns across temporally spaced trials in association with hit responses, but not during non-detections. Contrary to models relying on temporally stable networks or bulk-signaling, these results suggest that detection depends on transient differentiation in neuronal activity within cortical populations.

**Introduction**

Lesion studies in humans and animals indicate the causal importance of the primary visual cortex (V1) in detection, discrimination and awareness of visual stimuli (Lashley 1943; Weiskrantz et al. 1974; Weiskrantz 1996), and this role has been recently confirmed by direct optogenetic inhibition of mouse V1 (Glickfeld, Histed, and Maunsell 2013). Visual perception has been proposed to arise from interactions between stimulus-specific processing in V1 and neural activity in higher visual and frontoparietal areas, involving both feedforward propagation of activity and recurrent, top-down feedback (Shadlen and Newsome 1996; Britten and van Wezel 1998; Lamme et al. 2000; Haynes, Driver, and Rees 2005). Critical in unravelling neural correlates of vision is how detected and undetected stimuli are processed differently, especially when these stimuli are physically identical. For instance, it has been suggested that the intensity, duration and reproducibility of sensory neural activity may provide signatures critical for visual perception (e.g. Moutoussis and Zeki 2002; Schurger et al. 2010). In addition, it has been proposed that neural activity in V1 does not correlate with visual perception because stimuli that were seen or not seen evoked similar V1 blood-oxygenation level dependent (BOLD) signals (Vuilleumier et al. 2001; Rees et al. 2000), but this remains an area of substantial controversy (Ress and Heeger 2003; Palmer, Cheng, and Seidemann 2007; Nienborg and Cumming 2014). In this context it is important to recall that fMRI, EEG and MEG rely on a mean-field approach, leaving open the possibility that neural correlates of perception may be coded in more subtle ways that take into account the local differentiation present in populations of sensory neurons.

Such local, functional differentiation is supported by single- or multi-unit recording studies in visual, auditory and somatosensory areas of animals trained to make perceptual decisions (Logothetis, Pauls, and Poggio 1995; Britten et al. 1996; Posner and Gilbert 1999; Petersen, Panzeri, and Diamond 2002; Luna et al. 2005; Palmer, Cheng, and Seidemann 2007; Mitchell, Sundberg, and Reynolds 2009; Cohen and Maunsell 2009; Cohen and Maunsell 2011;

Sachidhanandam et al. 2013; Chen et al. 2013; Miyashita and Feldman 2013; Doron et al. 2014; McGinley, David, and McCormick 2015). Over the last decade it has become clear that the shared response variability between neurons (i.e., noise correlation) might be particularly important for sensory processing, because noise correlations can influence the amount of information that can be extracted from neuronal population codes (Averbeck, Latham, and Pouget 2006; Cafaro and Rieke 2010). Furthermore, it has been observed that these correlations can be reduced during stimulus presentation (Gutnisky and Dragoi 2008; Snyder et al. 2014) and directed attention, which may aid in disentangling stimulus information from noisy population responses (Mitchell, Sundberg, and Reynolds 2009; Cohen and Maunsell 2009; Herrero et al. 2013).

Although noise correlations have been studied well, they have the drawback of not being an instantaneous measure – their computation requires integrating neural activity over multiple time points or stimulus repetitions. Instantaneous aspects of population activity in cortex, such as temporal spike co-occurrence and population sparseness, seem critical for efficient neural coding (Olshausen and Field 1997; Vinje and Gallant 2000; Benucci, Saleem, and Carandini 2013; Luczak, Bartho, and Harris 2013). Some population-based measures have been proposed and tested in somatosensory and auditory cortex (Romo et al. 2003; Safaai et al. 2013; Carnevale et al. 2013; Buran, Trapp, and Sanes 2014). It has for example been shown that measures based on the variability and correlations between neurons correlate better with the animal's decision than simpler approaches based on the mean spiking rate (Safaai et al. 2013; Carnevale et al. 2013). However, in the domain of visual perception the behavioral relevance of only few population measures has been experimentally tested in paradigms where animals report behaviorally whether they have seen a stimulus or not.

Therefore we investigated correlates of visual stimulus detection using two-photon calcium imaging of populations of ~100 neurons in V1 L2/3 of mice performing a detection task, as superficial layers are easy to access with calcium imaging and have been reported to show neural correlates with stimulus detection (van der Togt et al. 2006; Ito and Gilbert 1999). Our first aim was to examine whether visual detection correlates with the mean visual response strength of V1 neurons, or rather with other metrics of population responses, such as noise correlation or variance. This led us to develop a novel population metric – response heterogeneity – that correlates better with stimulus detection performance, and particularly with the animal's reaction time, than traditional measures by capturing the dissimilarity of neuronal responses within a population. Second, an assumption in many computational models of vision is that neurons in distributed cortical architectures have relatively fixed roles in encoding visual features, but modulate their activation in a temporally dynamic manner based on attentional needs that can influence perception (e.g., Jones and Palmer 1987; Itti, Koch, and Niebur 1998; Desimone 1998; Dayan and Abbott 2001; Deco and Rolls 2004; Reynolds and Heeger 2009). To study whether modulations of neuronal activity that influence stimulus perception show temporally recurring patterns, we asked whether population activation patterns are more similar across trials that repeat the same stimulus presentation when the stimulus is successfully detected. We report that (i) visual stimulus detection does not correlate well with mean response strength, but is significantly correlated with population heterogeneity, (ii) neuronal populations show consistencies in activation patterns across temporally spaced trials in association with hit responses, but not when the animal fails to report a stimulus, and (iii) in addition to heterogeneity, multidimensional structures in neuronal population responses provide information on visual detection.

**Results**
To investigate how ensembles of primary visual cortex (V1) neurons are involved in visual detection we trained mice to perform a go/no-go stimulus detection task (fig. 1a). After task acquisition, we performed two-photon calcium imaging in V1 contralateral to the visually stimulated eye (fig. 1b,d; figure supplement 1-1). During imaging animals were awake, head-fixed and performed a detection task where they indicated by licking whether a square-wave drifting grating was presented. Stimulus duration was delimited by the onset of the first licking response, with a maximum of 3.0 seconds for no response; therefore no licks occurred during presentation of the stimulus. To acquire a sufficient range of hit/miss ratios we presented test stimuli with different luminance contrasts: 0.5%, 2%, 8% and 32%. These test trials were interleaved with 0% no-contrast and 100% full-contrast probe trials to estimate the animals' ratio of false alarms and omissions due to lack of motivation. For all analyses we discarded trials where animals responded within 150 ms after stimulus onset (0.3% - 3.5% of trials per animal), because such fast responses may be ascribed to spontaneous licking.

To quantify behavioral performance during execution of the task, we calculated the $2.5^{th}$ - $97.5^{th}$ percentile intervals (henceforth; 95% confidence intervals (CI)) of response proportions to the two types of probe trials: no-contrast and full-contrast stimuli. All eight animals (see Materials & Methods) showed a significantly above-chance visual detection of square-wave drifting gratings during the acquisition of neural data (fig. 1c) (non-overlapping Clopper-Pearson 95% CIs). Behavioral response proportions increased with higher stimulus contrasts (fig. 1e) (group-level linear regression analysis, $p<0.001$) and mean reaction times decreased (fig. 1f) ($p<0.005$).

**Response dissimilarity within neuronal populations correlates with detection**

As a first approach to examine population correlates of visual detection we investigated differences in mean activity levels between hit and miss trials (fig. 2). We defined each neuron's response during a trial as the mean dF/F0 during the entire stimulus presentation (fig. 2a,b). Because hit/miss would arguably be stronger in the population of neurons that prefer the features of the visual stimulus, we started out with investigating neural correlates of detection in the preferred population. We therefore calculated each neuron's preferred stimulus orientation (see Materials and Methods), and for the analysis in fig. 2c,d took for each trial the responses of only neurons that preferred the presented stimulus orientation (henceforth 'preferred population'). In fig. 2c, all trials of a single animal were grouped by stimulus contrast and behavioral response (hit/false-alarm ('response') or miss/correct-rejection ('no-response')), and the average preferred population response was calculated for hits and misses. As expected, the mean response increased with higher stimulus contrasts (fig. 2c,d and figure supplement 2-1 for traces across time). However, for this animal we did not find a significant difference between hit and miss trials for any individual contrast (false-discovery rate (FDR)-corrected paired t-test, $p>0.05$ for all contrasts) (note that for both false alarms and correct rejections V1 mean population response was indistinguishable from zero (fig. 2c,d, 0% contrast; figure supplement 2-1a)). When grouping the test contrasts (0.5% – 32%), the data did show a modestly higher response for hit than miss trials for single animals as well as across animals ($p<0.05$). We therefore asked whether this increase in neuronal responses during stimulus detection was due to consistent response enhancements of specific neurons, or due to a population-distributed process.

Including again also non-preferred neurons for all further analyses (unless stated otherwise), we calculated the hit modulation (dF/F0 increase during hits relative to misses) per neuron per hit trial (see Materials and Methods) (fig. 2e(I)) and investigated whether this hit-modulation could be explained by a subgroup of neurons that consistently enhances its activation during detection trials, random trial-by-trial population fluctuations, or both (fig. 2e). Hit-modulation was explained to a small, but significant extent by neuronal identity ($R^2=0.059$; $p<0.05$; fig. 2e(II),2f), and to a larger extent by population fluctuations across trials ($R^2=0.248$; fig. 2e(III)) or both processes together ($R^2=0.281$; fig. 2e(IV)). The number of consistently hit-modulated neurons (which could be either up- or downregulated) was significantly above-chance (fig. 2f; p<0.05). This pattern was robust over animals, as hit-modulations could be explained with above-chance accuracy by neuron identity, population fluctuations, or both in 7/8, 8/8 and 8/8 animals respectively (fig. 2g). The fraction of significantly hit-modulated neurons was above-chance (at p<0.05) for 7/8 animals. Although significant for most subjects, the variance explained by the consistency of neuronal responses was fairly low (always $R^2<0.1$), and even the combination of trial-by-trial population fluctuations and neuronal identity never exceeded $R^2=0.35$. This could indicate either that detection-related neural correlates in V1 are minor, or that a simple enhancement of mean activity is an index ill-suited to describe potentially strong, but more complex changes in neuronal population dynamics. In particular, we hypothesized that correlates of stimulus detection may be unfolding by multi-neuron interactions at the single trial level and rely on the relative contrast in activation between neurons.

Several metrics aim to quantify response heterogeneity within neuronal populations, such as the sparseness (Field 1994), or variance (Seung and Sompolinsky 1993). However, such metrics are rarely studied in the context of behavioral relevance and in the few cases where they are, their ability to predict behavior appeared modest (Froudarakis et al. 2014). Therefore we developed an alternative measure of population heterogeneity that aims to capture the spread in normalized population activity (fig. 3a,b; see also Materials and Methods): by subtracting the z-scored response (each trial being a single data point per neuron, see eq. 2) of each neuron from that of all other neurons in that same trial, we obtained a Δz-score matrix where high values indicate high pairwise dissimilarity in neuronal activation. Taking the mean over all pairwise Δz-scores provides a measure of population heterogeneity that can in theory be computed over an arbitrarily small time interval (but note that for all analyses, except those shown in fig. 5, we used a single trial as time unit). This way, similarly strongly activated as well as similarly weakly activated pairs of neurons will decrease heterogeneity. By contrast, dissimilarly activated neuronal pairs (i.e., one strong, one weak) will increase it. Therefore population heterogeneity incorporates both trial-by-trial fluctuations and intra-population differences in a neuronal pairwise manner. Its dependence on z-scored activity means that a neuron's contribution to heterogeneity is scaled to its relative level of activation - and because highly active neurons are often highly variable (Baddeley et al. 1997; Montijn, Vinck, and Pennartz 2014) also to its signal-to-noise ratio.

We applied this metric to the activity of neurons from the entire population during hit and miss trials and found a stronger correlation with behavioral stimulus detection than for mean response strength (see video 1 and fig. 3c for a single animal example). Test contrasts (0.5% – 32%) showed a highly significant overall increase in heterogeneity for hit trials (fig. 3c, paired t-test, $p<0.001$), but such modulations were absent for probe trials. This difference was consistent over animals (fig. 3d) and showed similar patterns for the within-preferred and within-non-preferred population heterogeneity (fig. 3e and figure supplement 3-1). Using a measure of effect size over animals (Cohen's *d*) we observed that heterogeneity showed a stronger correlation with visual detection than mean dF/F0 (fig. 3f; three paired t-tests vs. dF/F0 were all $p<0.05$). Linear single-trial prediction of hit or miss responses with a Receiver Operating Characteristic (ROC) analysis on either mean dF/F0 or heterogeneity showed that behavioral responses could be predicted above chance at single trial basis with both metrics, but heterogeneity showed a significantly higher prediction score (area under curve; AUC, t-test across animals, dF/F0 vs. 0.5; $p<0.05$, heterogeneity vs. 0.5; $p<0.001$, dF/F0 vs. heterogeneity; $p<0.01$) (fig. 3g,h). These results show that correlates of visual detection are better captured by the strength of pairwise response dissimilarities within the neuronal population than to overall increases in mean activation (but note that correlation-based measures also work well for hit-miss differentiation; figure supplement 3-2).

**Heterogeneity predicts reaction time**

Our observations suggest that not a general gain increase in population activity, but rather more complex changes in response strengths within a population determine the behavioral accuracy. Behavioral reaction time is often used as a proxy for salience, attention and readiness (Beck et al. 2008), so we hypothesized that similar dissociations for fast/slow responses may be found as for hit/miss trials. We performed linear regressions per animal for dF/F0 (fig. 4a) and heterogeneity (fig. 4b) as a function of reaction time. Similarly to hit/miss differences, the preferred population dF/F0 was not significantly associated with behavioral performance (regression slopes vs. 0, FDR-corrected one-sample t-test, n.s.), nor were preferred population z-scored activity, variance, sparseness, instantaneous Pearson-like correlations (see Materials and Methods), whole-population (raw and z-scored) dF/F0, and sliding-window based correlations (fig. 4c, figure supplement 4-1). However, heterogeneity and the spread in instantaneous Pearson-like correlations were inversely correlated with reaction time ($p<0.001$, $p<0.01$ respectively) and explained significantly more reaction-time dependent variance in the data than all other measures (FDR-corrected pairwise t-tests, heterogeneity vs. all, $p<0.05$). This relationship holds when analyzed over animals (fig. 4c) as well as per individual animal (figure supplement 4-1).

Our definition of heterogeneity is computationally somewhat similar to the width of the distribution of pairwise neuronal correlations in a population (see Materials and Methods). However, whereas the spread in instantaneous Pearson-like correlations is based on multiplying z-scored pairwise neuronal responses and taking their standard deviation, the heterogeneity metric (which instead uses the mean absolute distance in z-scored dF/F0 between pairs of neurons) was an even better predictor of behavioral reaction times (fig. 4c, $p<0.05$). As such, it is more closely related to the population mean of non-directional neuron-pairwise Mahalanobis (i.e., normalized Euclidian) distances than Pearson correlations. Our analysis shows that visual detection correlates well with large mean Mahalanobis distances in neural activity between pairs of neurons; i.e., a high heterogeneity in population activity.

Nonetheless, it could be argued that changes in population activity might be uncorrelated with the fidelity with which the population code represents visual stimuli. To address this, we used a Bayesian maximum-likelihood decoder to assess the presence of a stimulus from V1 population activity (see also Materials and Methods and (Montijn, Vinck, and Pennartz 2014)). Decoding performance was higher for behaviorally correct detection trials (fig. 4d; hit vs. miss, $p<0.05$) and the performance was similar to the animals' actual behavioral performance at a global level across contrasts (shuffled vs. non-shuffled, $p<0.001$; fig. 4e), as well as at a single trial level (chi-square similarity analysis of hit/miss trials for behavioral response and stimulus presence decoding, $\chi^2=135.36$, $p<10^{-30}$; figure supplement 4-2). Moreover, additional analyses revealed that stimulus features (orientation, contrast) were better decodable when the animal made a correct detection (figure supplement 4-2a) and when heterogeneity was high (figure supplement 4-2b-d). Thus stimulus features, such as orientation, are represented more accurately by neuronal populations in V1 during hit trials, even though the specific orientation was irrelevant for the animal to perform the visual stimulus detection task.

During higher levels of arousal, it has been observed that neuronal activation is more desynchronized (Cohen and Maunsell 2009; Froudarakis et al. 2014). Based on our current observations this led us to hypothesize that a high heterogeneity in V1 populations reflects a brain state conducive to stimulus detection. If correct, heterogeneity immediately prior to stimulus presentation should be predictive of reaction time. To test this, we split all hit trials into the slowest 50% and fastest 50% per contrast (see fig. 5a-d for examples) and calculated a measure of predictability of slow vs. fast responses based on the three seconds preceding stimulus presentation (fig. 5e; figure supplement 5-1). Using pre-stimulus-onset heterogeneity, fast response trials were highly predictable (FDR-corrected one-sample t-tests, p<0.01), while slow vs. miss trials were not (p=0.799). Behavioral responses were not predictable based on population dF/F0 (slow-miss, p=0.157; slow-fast, p=0.811; fast-miss, p=0.924), and the difference in predictability between heterogeneity and dF/F0 was significant for slow-fast (p<0.01) and fast-miss (p<0.05), but not for slow-miss (p=0.477) trials.

Although heterogeneity before stimulus onset thus predicts behavior, we also found a dissociation between detected (slow and fast responses) and undetected stimuli (miss trials) in the rise time latency to maximum heterogeneity upon stimulus onset (p<0.05). Detected trials correlated with fast rise times, while neuronal response heterogeneity to undetected stimuli ramped up much more slowly (fig. 5f). This argues against the interpretation that heterogeneity merely reflects a tonic brain state which can be fully gauged before stimulus onset. The formation of non-homogenous response patterns within neuronal populations is also related to the actual detection of visual stimuli, and constitutes a second effect in addition to background heterogeneity.

**Temporal consistency of the population code**
So far we have mainly addressed static differences in population activity structure correlating with behavioral responses. However, population codes can show complex temporal properties, such as transient formation of assemblies (Miller et al. 2014; Luczak, Bartho, and Harris 2013). After confirming the stability of our recordings to avoid potential confounds (figure supplement 1-1), we addressed whether such temporal population structures might offer additional insight in neural mechanisms of visual detection. We again split the data into miss, fast and slow response trials, and computed the correlations between response patterns from different trials separately for preferred and non-preferred neuronal populations (fig. 6a,b). Note that this analysis is not sensitive to potential non-stationary effects that might create artificial differences, because all stimulus types and behavioral responses are intermingled in time.

Within the preferred population, as well within the non-preferred population, we found that neuronal population activity patterns were more similar during fast trials, with a trend for slow trials, than during miss trials (preferred population, miss-slow; p=0.081, miss-fast; p<0.05, non-preferred population, miss-slow; p<0.05, miss-fast; p<0.05) (fig. 6c,d). To rule out that this effect might arise from biases in the analysis, we also compared these population pattern consistencies to those obtained from a shuffling procedure within stimulus types (see Materials and Methods). Similarly, population pattern correlations were significantly higher than shuffled for fast and slow trials, while miss trial consistency was not statistically different from the shuffled control (both preferred and non-preferred populations, paired t-tests shuffled vs. real, miss; p>0.3., slow and fast; p<0.05). However, note that these pattern consistencies are relatively low: they cannot fully account for the population activity structure and must therefore be interpreted as happening against a background of dynamic population activity.

**Analysis of heterogeneity in multidimensional space**

Most of our results so far have focused on the experimentally observed differences in hit-miss and reaction-time dependent effect sizes between heterogeneity and mean population activity, but have not addressed the question how this metric might be interpreted within a theoretical framework. Although heterogeneity correlates better with behavioral responses, and especially with reaction time, than many other metrics, this does not exclude that heterogeneity might be an epiphenomenon. We will address this issue next (see also Materials and Methods, section "Analysis of multidimensional inter-trial distance in neural activity") using an alternative definition of heterogeneity extended to multidimensional space. This alternative definition is required to study multidimensional properties of population responses, but yields a very similar correlation with stimulus detection as our pairwise definition of heterogeneity (figure supplement 3-2).

Neuronal population activity during any time epoch can be visualized as a single point in multidimensional neural space. For instance, the mean output in spikes per second of a "population" of two neurons will always be somewhere within a two-dimensional space bounded by the minimal and maximal neural activity of these two neurons (fig. 7a). Within a normalized version of this space a change in mean neural activity will always be parallel to the main diagonal that crosses the origin (minimal neural activity of both neurons) and the point of maximum activity for both neurons. This is true for populations consisting of two neurons, but can be readily extended to any number of dimensions (fig. 7b). Heterogeneity on the other hand does not change when a point moves across this diagonal (the difference will be zero regardless of whether all neurons are firing at 0 spikes per second, or their maximum), but rather changes as a point moves orthogonally to this diagonal (fig. 7c).

The observation we present in our current study - viz. that heterogeneity correlates with hit responses - can be explained by two mutually exclusive hypotheses; 1) the basis for hit and miss-related responses in V1 resides in specific regions in multidimensional neural response space (i.e., discrete states in the neural circuit) and therefore heterogeneity is an epiphenomenon (fig. 7a,b), or 2) neuronal response heterogeneity *per se* is important for stimulus detection. The latter implies that neuronal population response patterns during hit and miss trials should be distributed symmetrically around the main diagonal (which is the gradient along which the mean changes, as well as the axis where heterogeneity is zero). This is because regardless of the specific location in multidimensional space, heterogeneity only captures the distance to this diagonal; rotation or mirroring around this diagonal therefore does not change heterogeneity, but does in fact change the distribution in multidimensional space (fig. 7c,d).

Accurately quantifying a distribution's shape in multidimensional space requires exponentially more data points as the number of dimensions increases. For direct quantification, our current data set is unfortunately insufficiently large. Therefore, in order to estimate multidimensional symmetry around the main diagonal, we decided to study the effect of mirroring points across this diagonal. First, we calculated the distribution of pairwise inter-point distances in neural response space without mirroring (fig. 7e). In this case, each point again represents the population response during a single trial. The data show that the inter-point distance was slightly, but significantly larger for hit than miss trials (paired t-test, $p<0.05$) (fig. 7f). This indicates that population responses during hits encompass a larger volume of neural response space than during misses, which will increase heterogeneity values and allows more information to be encoded with the same number of neurons.

To assess symmetry specifically, we mirrored each trial one at a time across the main diagonal and recomputed in each case the distribution of pairwise inter-point distances. If the population responses are distributed asymmetrically around the diagonal, then mirroring will increase the pairwise distance, while if they are distributed symmetrically no change should be observed. There was a small, but significant increase in inter-point distance for both hits and misses, and mirroring increased the inter-point distance more for hits than misses ($p<0.05$; fig. 7f, inset). Although the effect sizes are small, they are significant, and we can therefore – at least based on this analysis - not (yet) conclude that heterogeneity is more than an epiphenomenon. In fact, the difference between hits and misses suggests that population responses during hit trials are more asymmetrical (i.e., more clustered in discrete states of neural activity) than during miss trials (fig. 7f, inset). Considering that intertrial population pattern consistencies were lower during miss trials (fig. 6) we can conclude that neuronal populations during miss trials show more random behavior within a limited neural space, while neuronal populations during hit trials show more structured behavior in a more extended neural space.

Noting the small effect size of the previous analysis, we asked whether the removal of the mean of the neuronal population response was as detrimental to encoding hit/miss differences as the removal of heterogeneity. Again visualizing population responses as points in neural space, one can remove any differences in mean response between trials by projecting all population responses to a plane orthogonal to the diagonal, or remove any differences in heterogeneity by projecting all points onto a manifold at a fixed distance from the diagonal (see Materials & Methods, and fig. 7g). To test the effect of these removals on hit/miss differences, we performed a decoding procedure of hit vs. miss trials (i.e., we decoded the animal's response) on the original data, and on data with the mean, the heterogeneity or both aspects removed. The results show no difference between the original data and the mean-removed data, but removing heterogeneity (or both heterogeneity and the mean) led to a small, but significant decrease in decoding performance (heterogeneity-removed vs. original data, $p<0.05$; heterogeneity-removed vs. mean-removed, $p<0.05$) (fig. 7h). However, even with heterogeneity removed, decoding performance was still well above chance (63% correct for original data, 60% correct for both removed; 50% is chance).

We therefore conclude that heterogeneity contributes significantly as a non-epiphenomal population property to the differentiation in neural responses between visual detections and failures to detect a stimulus, but also that most information resides in neuronal population response patterns other than its mean or heterogeneity. Moreover, the mean response of a neuronal population is less important than its heterogeneity. While neural response heterogeneity may be an important factor and useful metric for its strong correlation with especially reaction times, further research is required to discover which other neural properties may be important for visual stimulus detection.

**Discussion**

We found that behavioral stimulus detection correlates more with non-linear neuronal population activation patterns, such as heterogeneity, correlations and variance, rather than overall response strength in L2/3 of mouse V1 (fig. 2,3; figure supplements 2-1, 3-1, 3-2). Using a novel measure of population heterogeneity we show that the differentiation in activation within these populations predicts visual detection, and particularly behavioral reaction time, and is associated with an increased accuracy of stimulus presence and feature representation by the population (fig. 4; figure supplement 4-2). High heterogeneity prior to stimuli correlated with fast hit responses, but also showed a dissociation between detection and non-detection behavior, indicating that detection-related population activity may be gated by arousal mechanisms (fig. 5; figure supplement 5-1). Neuronal population activation patterns are more similar during accurate task performance upon repetition of the same stimulus, but not when the animal fails to respond, suggesting that specific population patterns may recur when the animal is well engaged in the task (fig. 6). Taken together, these results suggest that neural processing of information related to detection behavior depends on transient differentiation in neuronal activity within cortical populations, rather than on temporally stable ensembles or on gain-modulation of population activity as a whole (fig. 7).

**Potential confounds**

Our analyses show differences between hit and miss trials that we interpret as being related to perception and visual processing. However, in principle the observed differences could be subject to a number of confounds that might limit their interpretability. First, the relatively mild water restriction led to a behavioral performance at the 100% probe trials that is lower compared to other studies using similar tasks (Glickfeld, Histed, and Maunsell 2013). This could mean that the observed differences in heterogeneity between hit and miss trials are due to changes in motivation, rather than visual detection. If this were the case, however, hit-miss differences should be as large during 100% probe trials as during test contrast trials. Figure 3 and figure supplement 2-1f shows that this is not the case; during intermediate test contrasts (2% and 8%) the hit-miss difference in heterogeneity is largest. A similar reasoning applies to 0% contrast probe trials, where a heterogeneity difference between (false alarm) responses and correct rejections was lacking, which strongly argues against heterogeneity being due to response emissions per se. Thus, it is highly unlikely that the neural correlates of behavioral responses as reported in this study are due to differences in motivation. Given this caveat on suboptimal performance, one may predict that a better behavioral performance would have only increased the hit-miss effect sizes we report in this study.

Other potential confounds include instabilities in z-plane focus, other locomotor-related artifacts and running-induced modulations, as it has been reported that behavioral activity and running can induce instabilities in the plane of focus with awake two-photon calcium recordings, as well as changes in the neuronal responses in mouse V1 (Dombeck et al. 2007; Niell and Stryker 2010; Saleem et al. 2013). To address potential z-shifts during the acquisition of neural data we compared each imaging frame to 3D anatomical z-stacks acquired after recording the neural data (see Materials and Methods). Slight changes in z-location were detected after the onset of hit responses by licking, but these were mostly confined to the reward presentation period (which was not used for any of our analyses) and rarely exceeded more than a couple of microns (figure supplement 1-1). Moreover, exclusion of trials where mice were running did not qualitatively change hit-miss differences in neuronal activity (figure supplement 2-1g,h), nor did using only the first 400 ms after stimulus onset to avoid licking-preparation feedback from other brain regions (figure supplement 2-1i,j).

To control for potential confounds related to eye movements and blinking, we analyzed eye-tracking videos to detect blinks and saccades. After removing all trials where the animals were making saccades or were blinking, we again found no qualitative difference with the results we observed previously (figure supplement 2-1k,l), but did observe a small, but significant correlation between pupil size and heterogeneity, suggesting higher heterogeneity with increased arousal states (figure supplement 4-1e). Overall, we conclude that the neural correlates we report here, and interpret as related to perception, are most likely not due to recording instability, changes in motivation, locomotion, motor-related signals associated with licking, or eye movements and blinking.

**Possible cortical layer specificity of poor correlation of mean population responses**

Importantly, our results only pertain to L2/3 of the primary visual cortex in mouse, which does not exclude the possibility that the mean population response of for example deeper layers (L5) in V1 would correlate better with visual stimulus detection. Previous research has shown that extensive differences exist between superficial and deep layers: whereas L2/3 neurons often show relatively low peak firing rates and sparse responses to sensory stimulation, L5 neurons show denser response patterns with on average higher peak firing rates (De Kock and Sakmann 2008; Harris and Mrsic-Flogel 2013). In somatosensory cortex it has been shown that hits and correct rejections in a Go-NoGo object localization task can be better separated using mean spiking rates in L5 than in L2/3 (O'Connor et al. 2010). Our result that L2/3 populations show only a small differentiation between hit and miss responses in mean activation should therefore not be taken as proof for a canonical principle also applicable to other cortical layers. Future validation of our results in deep layers is necessary for a decisive answer whether our results are indeed applicable to different layers of primary sensory cortex.

**Comparison with other studies and neural interpretation of heterogeneity**

Our task design included drifting gratings with different orientations, with the qualification that orientation was task-irrelevant for the mice, as they were only required to detect stimuli whenever they appeared. Our observation that stimulus features are represented more accurately (as quantified by decoding accuracy) during hit than miss trials may therefore be somewhat surprising (figure supplement 4-2a). This suggests that mechanisms that increase the likelihood of stimulus detection may be acting through a general enhancement of stimulus processing intensity, corroborating previous research in monkey showing that attention can lead to horizontal shifts in contrast response curves, as if the stimulus were of higher contrast (Martínez-Trujillo and Treue 2002). It is interesting to ask whether our results on heterogeneity can be cast in terms of dynamic range effects. Neurons are expected to climb in this dynamic range when visual contrast increases, which is confirmed by the rise in dF/F0 (fig. 2d). However, if heterogeneity would be primarily determined by neurons being able to operate along the steep slope of their dynamic range, then the large difference in heterogeneity between hits and misses (fig. 3c) along the test contrasts (0.5-32%) would not be expected.

Of further interest is to compare our results on heterogeneity with studies reporting that sparseness in L2/3 populations of rodent V1 is high during passive viewing (Barth and Poulet 2012), depends on cortical state and improves neural discriminability during passive processing of natural scenes (Froudarakis et al. 2014). Although in our analysis sparseness and variance explained more behavioral variability in reaction time than (z-scored) mean population activity (fig. 4c), these measures perform much worse than heterogeneity and the spread of instantaneous Pearson-like correlations. Possibly a sample of ~60-70 tuned neurons is insufficient to estimate instantaneous sparseness accurately. An alternative explanation for this poor correlation could be that sparseness of L2/3 populations results from anatomical wiring required for efficient stimulus coding and to enable locally selective synaptic plasticity without immediately changing the coding of stimulus features within the population response (Rao and Ballard 1999). Correlates of visual detection that depend on accurate stimulus feature representation might then be better captured by a maximization of Mahalanobis distances in neural activity between pairs of neurons within this already sparse network. This latter interpretation is in line with the data we recorded and suggests that sparse stimulus representation by L2/3 neurons reflects a structural optimization of the population code to represent stimulus features, while heterogeneity captures more temporally dynamic modulations related to perception.

**Multidimensional analysis: heterogeneity contributes to, but does not fully account for visual detection**

In addition to our approach based on pairwise relations in neuronal responses, we investigated multidimensional patterns of population activity (fig. 7). These results indicate that, while heterogeneity is more important for separating stimulus detections from non-detection in neural response space than the population mean, these properties combined still cannot capture the full set of neuronal response characteristics that define the accurate detection of visual stimuli in L2/3 of mouse V1. This suggests that other patterns of population activity, such as potentially transient assembly formation, may be important for visual stimuli to be correctly detected. From our multidimensional analyses we can conclude that simple bulk-approaches (i.e., correlating a population's mean response with behavioral output) are insufficient when one aims to address how early sensory cortex areas are involved in the processing and detection of visual stimuli.

Related to these findings is our observation of behavioral-state specific consistencies in population activation patterns across trials. This provides some constraints on how population heterogeneity is modulated at a neurophysiological level. Neuromodulators such as acetylcholine (ACh) and noradrenaline are correlated with attention and arousal, and may influence cortical population dynamics (Metherate, Cox, and Ashe 1992; Coull et al. 2004; Pinto et al. 2013), such that they facilitate repeated activation of similar subnetworks of neurons within a population responding to the same stimulus. Without such neuromodulators neurons within the same preferred population would randomly participate in representing the current stimulus. This interpretation is compatible with previous work; for instance, ACh has been observed to influence burst spiking, membrane potential fluctuations, cortical oscillations and desynchronization. These processes have been implicated in modulating competitive inhibition effects within neuronal populations and may very well influence the consistency of specific neuronal subnetworks being activated (Börgers, Epstein, and Kopell 2008; Fries 2009; Bosman, Lansink, and Pennartz 2014). If heterogeneity in a recurrently connected V1 population is in part determined by suppression of the most weakly by the most strongly stimulus-driven neurons, then behaviorally correlated heterogeneity enhancements may be another facet of arousal as well as perception-related modulations of stimulus-evoked population activity.

Population coding phenomena have long been hypothesized to be important for sensory processing, but so far few studies have investigated their relevance for perceptual decisions. Here, we show that population heterogeneity is correlated with behavioral stimulus detection and that it predicts correct behavioral performance. Our results imply that neurophysiological measures dependent on population averages (i.e., multi-unit activity (MUA), electro-encephalograms (EEG) and functional magnetic resonance imaging (fMRI)) may underestimate the correlation between visual detection and V1 L2/3 activity because the assumption of population response homogeneity is violated especially during active processing of visual information. In short, our results support contrast-sensitive changes in mean population activity during visual task performance (fig. 3c,d), but stress the importance of population recordings with single-cell resolution (fig. 4c-f).

**Materials and Methods**

**Animals and surgery**

All experimental procedures were conducted with approval of the animal ethics committee of the University of Amsterdam (cf. (Goltstein et al. 2013; Montijn, Vinck, and Pennartz 2014)). Experiments were performed on eight adult, male wild-type C57BL/6 mice (Harlan), 128-164 days old at the day of calcium imaging (29.1 – 32.7 grams). Prior to the imaging experiment, all animals were surgically fitted with a head-bar implant and trained head-fixed for up to three months to perform a visual go/no-go detection task. At the day of the imaging experiment, we performed intrinsic signal imaging to define the area corresponding to the retinotopic region in V1 responsive to the visual stimulus. We performed a small (1.5-2.0mm) craniotomy at that location and used multi-cell bolus loading with Oregon Green BAPTA-1 AM to record calcium transients and Sulforhodamine-101 to label astrocytes (Stosiek et al. 2003; Nimmerjahn et al. 2004).

**Behavioral training**

Mice were trained five days per week, each for approximately 45 minutes per day, on a head-fixed go/no-go visual detection task over a period of 10-12 weeks, where we aimed to get sufficient hit as well as miss trials for test contrasts. Mice were water-deprived for 6 hours preceding training and otherwise had ad libitum access to water. Weight was monitored 3 times per week and never dropped below 90% of their non-restricted growth curve. Behavioral training was performed inside four dark, sound attenuated chambers and occurred during the active (dark) cycle of the animals; each animal was always trained in the same behavioral setup. We did not observe any deviant learning effects associated with any specific behavioral setup (data not shown). During the first 5 days of training we conditioned licking in response to visual stimulation by pairing passive stimulation with reward delivery (~9 microliter of water with 15% sucrose with 1% vanilla extract) (stage 1). After the conditioning phase, visual stimuli (100% contrast drifting gratings as described in previous paragraph) were presented indefinitely until mice made a licking response that was monitored using a custom-built infrared LED based lick detector. When animals made a response, the visual stimulus presentation terminated and reward was available for 5 seconds. This shaping phase (stage 2) lasted for a maximum of 5 days or less if the animals were often making clear lick responses. After this ~2 week initial phase we started training the animals on a simple version of the final task (stage 3); maximum stimulus presentation was reduced to 5 seconds and subsequent trials would only start if the mice did not make any lick responses for at least a random interval of 1-3 seconds. During this stage reward size was gradually reduced to ~3 microliters per trial. When animals would consistently perform at least 80 trials within a period of 45 minutes, they would be moved to the next stage. In stage 4 we introduced 0% contrast probe trials to monitor the behavioral performance of animals by testing for false-alarm responses and calculating if they showed statistically significant above-chance performance. In this stage we also lengthened the inter-trial interval to any random duration between 6-8 seconds. Once mice attained a sufficient ratio of hit/miss trials we moved them to training stage 5, where we increased the inter-trial interval to 10-12 seconds and presented mild air puffs as a negative reinforcer whenever mice would lick outside the stimulus presentation or reward delivery period. At this stage, animals were required to not lick for a random interval of 1-3 seconds in order to gain access to the next stimulus presentation. Stage 5 lasted until the mice had been trained for 8-10 weeks in total. Finally, if mice performed consistently and significantly above chance during stage 5 (n = 12 / 21 animals), then in the two-week period preceding the imaging experiment mice were trained on the microscope setup, and our setup's resonant mirrors were activated to produce the characteristic 8000 Hz sound that would also be present during calcium imaging. In this final stage all possible efforts were made to simulate surroundings of the eventual calcium imaging experiment as closely as possible to habituate the mice to the two-photon laser lab's environment. Mice were always allowed to take up to three seconds after stimulus onset to respond and were thus not explicitly trained to make fast behavioral responses.

**Craniotomy and dye injection**

On the day of the two-photon calcium imaging experiment, buprenorphine (0.05mg/kg) was injected subcutaneously 30-60 minutes before induction of anaesthesia with isoflurane (4.0% induction, 0.8% maintenance during intrinsic signal imaging, 1.5-2.5% maintenance during invasive surgical procedures). After induction, the animal was placed in a custom-built head-bar holder designed for performing surgical procedures. We removed the cover glass, silicon elastomer and layer of glue covering the skull in the cranial window before performing Intrinsic Signal Imaging (ISI) to localize the precise location of our stimulus' receptive field location in the primary visual cortex (V1). We subsequently performed a small (1.5-2 mm) craniotomy above the retinotopic area responding to visual stimulation with drifting gratings. After the craniotomy, the dura was kept wet with an artificial cerebrospinal fluid (ACSF: NaCl 125 mM, KCl 5.0 mM, $MgSO_4 * 7 H_2O$ 2.0 mM, $NaH_2PO_4$ 2.0 mM, $CaCl_2 * 2 H_2O$ 2.5 mM, glucose 10 mM) buffered with HEPES (10 mM, adjusted to pH 7.4). After making the craniotomy, multi-cell bolus loading with Oregon Green BAPTA-1 AM (OGB) and Sulforhodamine 101 (SR101) was performed 230-270 microns below the dura as previously described in Montijn et al. (2014) and Goltstein et al. (2013). After injection of the dyes, the exposed dura was covered with agarose (1.5% in ACSF) and sealed with a circular cover glass that was fixed to the skull using cyanoacrylate glue. The animal was allowed to recover for a minimum of 90 minutes before starting the behavioral task and two-photon calcium imaging. Of the 12 mice that learned the task, 2 animals were rejected due to insufficient imaging quality.

**Visual stimulation**

All visual stimulation was performed on a 15 inch TFT screen with a refresh rate of 60Hz positioned at 16 cm from the mouse's eye, which was controlled by MATLAB using the PsychToolbox extension (Brainard 1997; Pelli 1997). Stimuli consisted of sequences of eight different directions of square-wave drifting gratings that were monocularly presented in randomized order. Visual stimulus duration started at infinite during the initial training phase and was gradually reduced to a maximum duration of three seconds for the final task stage. Stimuli were alternated by a blank inter-trial interval of variable duration (random minimum of 10-12 seconds) during which an isoluminant grey screen was presented. Visual drifting gratings (diameter 60 retinal degrees, spatial frequency 0.05 cycles/degree, temporal frequency 1 Hz) were presented within a circular cosine-ramped window to avoid edge effects at the border of the circular window. A field-programmable gate array (FPGA, OpalKelly XEM6001) was connected to the microscope and behavioral setup and interfaced with the visual stimulus presentation computer to synchronize the timing of visual stimulation with the microscope frame acquisition and behavioral setups.

**Z-drift quantification and recording stability analysis**

Slow z-drifts were quantified by comparing the similarity of 100 frames in the beginning, middle and end of each stimulus repetition set to slices recorded at different cortical depths (step size ~1-2 microns) before or after functional calcium imaging was performed for 5 of 8 animals. If z-drifts larger than 10 microns occurred slowly over multiple repetition blocks, or if slow z-drift was detected manually, the entire recording of a single animal was split into multiple analysis periods (n=2 populations for animals 1 and 7; n=1 population for all other animals) and analyzed independently (figure supplement 1-1). For the two animals for which we split the recordings, we afterwards averaged all measures over the two populations, yielding a single independent data point also for these animals for each measure.

To confirm the stability of our recordings, we performed a further analysis quantifying the discriminability of neurons relative to their surroundings over time (figure supplement 1-1). Therefore we calculated during each imaging frame the mean fluorescence of the pixels within the neuron's soma ($F_{soma}$) and the fluorescence of a neuropil annulus surrounding the soma ($F_{neuropil}$), which we defined as all pixels within a concentric band from 2 – 5 microns away from the soma. For each frame we then calculated the discriminability ratio $D_r$ as $D_r = F_{soma} / (F_{soma} + F_{neuropil})$, and set a threshold at $D_r=0.5$ (equal luminance of soma and neuropil). Whenever this measure dropped below the threshold, we calculated the duration of this epoch until it would return to above the threshold, and took the maximum duration of all these epochs as a single measure per neuron. Most neurons from all sessions showed maximum below-threshold durations near 0 seconds, and no neurons showed durations longer than one second (figure supplement 1-1).

To address the potential confound of fast changes in z-plane due to anticipatory fidgeting behavior by the animals, we calculated the depth of each imaging frame and analyzed whether responses to visual stimuli were preceded by shifts in z-plane that could influence our results. As can be seen in figure supplement 1-1l-o, z-shifts were mostly confined to the epoch immediately following hit responses, which are not used in our analyses, and in general z-shifts were very small and rarely exceeded more than 1 micron.

**Eye tracking**

We recorded eye movements during the entirety of the calcium imaging experiment to be able to correct for possible contamination of our results by excessive blinking and/or saccades. For this purpose we placed a near-infrared light sensitive camera (JAI CV-A50IR-C Monochrome 1/2" IT CCD Camera) with a large-aperture narrow-field lens (50 mm EFL, f/2.8) above the visual stimulation screen directed at the mouse's visually stimulated eye. Images were acquired at 25 Hz and pupil tracking was performed offline using custom-written MATLAB scripts. Eye position was used to control for possible saccade effects (figure supplement 2-1k,l), and pupil diameter was used to assess its correlation with heterogeneity (figure supplement 4-2e).

**Calcium imaging recordings and final task parameters**

Dual-channel two-photon imaging recordings (filtered at 500-550nm for OGB and 565-605nm for SR101; see fig. 1b) with a 512 x 512 pixel frame size were performed at a sampling frequency of 25.4 Hz. We used an *in vivo* two-photon laser scanning microscopy setup (modified Leica SP5 confocal system) with a Spectra-Physics Mai-Tai HP laser set at a wavelength of 810 nm to simultaneously excite OGB and SR101 molecules, as previously described (Montijn, Vinck, and Pennartz 2014) in cortical layer 2/3 at depths from the pia mater ranging from 140 to 170 microns (figure supplement 1-1, video 1). During data acquisition mice were performing a go/no-go stimulus detection task where the animals had to lick whenever a visual stimulus was presented. Stimulus parameters were equal to those described above. We varied the contrast of the drifting grating (0%, 0.5%, 2%, 8%, 32% and 100%) to elicit a wide range of hit/miss ratios. Responses to 0% contrast probe trials were not rewarded, but responses to all other contrasts were. We did not explicitly aim for very high detection performance (high hit rates and low miss rates) to avoid overtraining and associated habitual or automated responding (Balleine and Dickinson 1998). A complete set of visual stimuli therefore consisted of 48 trials (6 contrasts times 8 directions). The order of presentation of these 48 trials was randomized independently for each repetition block. After the experiment was completed we tested for statistically significant stimulus detection performance by calculating the binomial 2.5$^{th}$ - 97.5$^{th}$ percentile intervals (henceforth 95% confidence interval (CI)) of response proportion to the two probe trial types – 100% and 0% contrast stimuli – using the Clopper-Pearson (CP) method. Of the 10 animals from which we recorded calcium imaging data during task performance, one was rejected because of excessive variability in responses due to brain movement, and one was rejected due to insufficient discriminability between the two types of probe trials (overlapping CIs). All data we present in this paper are from the remaining eight animals. The number of repetitions per stimulus type (unique orientation x contrast) ranged from 6 to 16. For most analyses, we took the mean over all orientations (n=4), so each contrast was presented 24 – 64 times. For all analyses of single-animal data each trial was taken as a single data point, where its value was the mean dF/F0 over all recorded frames during stimulus presentation (which was dependent on the reaction time of the mouse). To avoid the confound of having higher signal-to-noise ratios for miss than hit trials due to longer data acquisition, within each contrast group we randomly assigned to all miss trials a duration randomly selected from the reaction time distribution of hit trials.

**Data preprocessing**

After a recording was completed small x-y drifts were corrected offline with an image registration algorithm (Guizar-Sicairos, Thurman, and Fienup 2008). To retrieve *dF/F$_0$* values from the recordings, regions of interest (ROIs; neurons, astrocytes and blood vessels) were determined semi-automatically using custom-made MATLAB software for each repetition block separately (see https://github.com/JorritMontijn/Preprocessing_Toolbox). For these ROIs we subsequently calculated *dF/F$_0$* values as previously described (Montijn, Vinck, and Pennartz 2014): For each image frame $i$ a single $dF_i/F_{0i}$ value was obtained for each neuron by calculating the baseline fluorescence ($F_{0i}$), taken as the mean of the lowest 50% during a 30-second window surrounding image frame $i$. $dF_i$ is defined as the difference between the fluorescence for that neuron in the given frame and the sliding baseline fluorescence ($dF_i = F_i - F_{0i}$) (Montijn, Vinck, and Pennartz 2014). The mean number of simultaneously recorded neurons/session was 92.6 (range: 68 – 130 (sd: 19.0) neurons). After this initial analysis all neurons were tested on consistency for preferred stimulus orientation and any neurons that showed inconsistencies over different repetition blocks (i.e., more than 1/3[rd] showing different preferred orientations) were rejected from further analysis (mean number of consistently tuned neurons per animal was 66.3 +/- 18.6 (70.8% +/- 7.75% of all neurons) (mean +/- sd) ). Unless otherwise specified, all analyses shown in this paper are based on across-animal meta statistics based on a set of 8 independent data points (one data point / animal) and all multiple comparison t-test p-values were adjusted by the Benjamini & Hochberg False-Discovery Rate correction procedure and were deemed significant if the resultant p-value was lower than 0.05. For quantification and control procedures related to z-drift and recording stability, see figure supplement 1-1. For control analyses where we performed neuropil fluorescence subtraction (figure supplement 4-2i), we used similar procedures as described previously (Greenberg, Houweling, and Kerr 2008; Mittmann et al. 2011); we calculated the correlation (r) between each neuron's somatic fluorescence and surrounding neuropil (annulus between two and five microns from soma) and corrected on each frame the neuron's fluorescence as follows: $F_{corr} = F_{soma} - r * F_{neuropil}$. Estimated neuropil contamination varied widely between neurons, but was generally in the range between 0.1-0.6, similar to previously reported values (Greenberg, Houweling, and Kerr 2008; Mittmann et al. 2011). We recomputed the explained variance of several metrics as a function of reaction-time (see fig. 4c, figure supplement 4-1) and found that neuropil correction did not affect our main conclusions (figure supplement 4-2i).

**Linear regression analysis**

All linear regressions were performed on single-animal data sets, yielding regression coefficients for the intercept and slope through minimizing the error between a linear function and the single animal's data points. Statistical significance was quantified by performing a one-sample t-test of the coefficients from all animals (n=8). Significance level was set at an alpha of 0.05 and p-values were adjusted if necessary by a post-hoc Bonferroni-Holmes correction.

**Calculation of preferred stimulus orientation**

We presented eight directions of visual drifting gratings and calculated the preferred stimulus orientation of all neurons by summing opposite directions as belonging to the same stimulus type, because the vast majority of mouse V1 neurons is tuned sharply to an axis of movement, but much less so to a specific direction within that axis (i.e., most neurons are strongly orientation-tuned, but less direction-tuned (e.g., (Andermann et al. 2011)). For these four orientations we took each neuron's mean response over all trials, and defined its preferred orientation as the stimulus that caused the highest mean dF/F0 value. For most analyses we used the neuronal responses to all orientations, except for fig. 2c,d, where we used only the response of the preferred orientation, as we hypothesized the preferred population might yield stronger hit/miss differences in neuronal activity.

**Predictability of hit-modulation by consistent neuronal responses and whole-population fluctuations**

To investigate the source of hit-related increases in population dF/F0 and determine if there might exist a subgroup of neurons that consistently enhances its activation during detection trials (as compared to non-detection), we defined a dF/F0 hit modulation index $\Psi$ for each hit trial ($t$) for each neuron ($i$) as the neuron's dF/F0 activity ($R$) relative to the mean ($\mu$) and standard deviation ($\sigma$) of its response during miss trials ($m$) of the same type (identical orientation ($\theta$) and contrast ($c$)):

$$\Psi_{i,t} = (R_{i,t} - \mu_{m,c,\theta}) / \sigma_{m,c,\theta} \qquad \text{(eq. 1)}$$

In other words, $\Psi_{i,t}$ of a given trial represents the z-scored dF/F0 activity relative to the neuron's response to the same stimulus when the stimulus remained undetected (fig. 2e, left panel). The hit-modulation matrix $\mathbf{\Psi}$ of all hit trials and all neurons can then be approximated by neuron identity (mean over trials), trial-by-trial fluctuations (mean over neurons), or both (addition of the matrices yielded by the two previous approximations) (fig. 2e). We then calculated the explained variance ($R^2$) of the population response pattern by its canonical equation based on the residual ($SS_{res}$) and total sum of squares ($SS_{tot}$). We defined $SS_{tot}$ as the sum of all squared values in $\mathbf{\Psi}$, and $SS_{res}$ as the sum of the squared differences between $\mathbf{\Psi}$ and the approximation matrix as defined above (by neurons, trials or both). To assess significance we performed 1000 shuffle iterations where we randomized neuronal identities per trial (for approximation by neuron identity), randomized trial identities per neuron (for approximation by trial identity), or randomized both (for approximation by both). Per shuffle iteration we calculated the explained variance, which yielded a shuffled distribution per prediction (e.g., fig. 2f). A prediction was defined as significantly above-chance when the real explained variance was at least two standard deviations away from the shuffled distribution mean (corresponding to p<0.05).

**Heterogeneity calculation**

We calculated heterogeneity of population activity as follows (see also fig. 3b). For each independent data source $i$ (i.e., a neuron) that provides a certain measurement $R$ at each time point $t$ (i.e., $dF/F_0$ activity of a single trial), we first z-scored the responses of $i$ over all trials $T$ (i.e., all contrasts and orientations). For all analyses we took $t$ to be a single trial, except those shown in fig. 5, where $t$ corresponds to a data acquisition point (i.e., a single calcium imaging frame), and calculated heterogeneity as follows. First, we z-scored all trial responses per neuron over all trial types (therefore high-contrast, preferred orientation stimuli yield higher z-score values than low-contrast, non-preferred orientations):

$$Z_{i,t} = (R_{i,t} - \mu_i)/\sigma_i \qquad (eq.\ 2)$$

**Z** is therefore a matrix containing $n$ (number of neurons) by $T$ (trials) measurements of standard deviations ($\sigma$) from the mean over all trials ($\mu$). Next, for each trial $t$ we calculated the pairwise distance (in standard deviations) from each independent source to each other independent source (pairwise neuronal $\Delta\sigma$): we repeated the z-scored population response vector $\mathbf{z}_t$ over its singular dimension $n$ times, where $n$ is the number of neurons in $\mathbf{z}_t$ (yielding a square matrix), subtracted this matrix from its own transpose $\mathbf{z}_t^T$, and took the absolute of the result, giving the heterogeneity matrix $\mathbf{H}_t$:

$$\mathbf{H}_t = |\ \mathbf{z}_t - \mathbf{z}_t^T\ | \qquad (eq.\ 3)$$

To get a single measure of population heterogeneity per trial ($h_t$), we next took the mean of all z-scored distances between neuronal pairs ($i,j$) in the heterogeneity matrix; this provides a measure of the mean distance in activation levels within our population at a single trial $t$:

$$h_t = \Sigma_{i=[1\ \ldots\ n-1]} \Sigma_{j=[i+1\ \ldots\ n]} (H_{t,i,j})/ ((n \cdot (n - 1))/2) \qquad (eq.\ 4)$$

**Effect size of mean population dF/F0 and heterogeneity**

We used a measure of effect size using Cohen's *d* to quantify which metric (mean dF/F0 or heterogeneity) showed a stronger correlation with visual detection. We calculated for both metrics per animal the effect size for all intermediate contrasts (0.5% − 32%) between hit and miss trials and took the mean over these four values, yielding a mean hit/miss effect size for dF/F0 and heterogeneity per animal. This allowed us to perform a paired t-test between the dF/F0 effect sizes and heterogeneity effect sizes to test for statistical significance. Cohen's *d* is defined as the difference between the two means (hit; $\mu_h$, miss; $\mu_m$) divided by the pooled standard deviation for the data;

$$d = (\mu_h - \mu_m) / \sigma_p, \qquad (eq.\ 5)$$

where $\sigma_p$ is defined as

$$\sigma_p = \sqrt{[(n_h - 1) \cdot var_h + (n_m - 1) \cdot var_m] / (n_h + n_m - 2)} \qquad (eq.\ 6)$$

**Instantaneous Pearson-like correlations & sliding window Pearson correlations**

For a pair of neurons *x* and *y*, the Pearson correlation (*R*) of their activity can be calculated by z-scoring each neuron's response vector (as in eq. 2) and taking the mean of the element-wise multiplication of the two vectors:

$$R_{x,y} = \Sigma_{t=[1\ \ldots\ T]} (Z_{x,t} \cdot Z_{y,t}) / T \qquad (eq.\ 7)$$

Here, notations are the same as for eq. 3-5; *t* is a single trial and *T* is the total number of trials. Using this equation, it is impossible to obtain an instantaneous correlation value between two neurons for each trial, because its calculation requires taking the mean over all trials. This poses a problem if we want to estimate the instantaneous correlation value between a pair of neurons for a given trial. Therefore we computed a modified measure, the instantaneous Pearson-like correlation ($\check{R}$). For each pair of neurons we calculated the z-scored element-wise product (each element being a single trial), which yields a three-dimensional matrix $\check{Z}$ with size [$n$ by $n$ by $T$], where $n$ is the number of neurons:

$$\check{Z}_{x,y,t} = Z_{x,t} \cdot Z_{y,t} \tag{eq. 8}$$

Taking the mean over the matrix's third dimension (trials) gives the conventional Pearson's pairwise correlation matrix over neuronal pairs. However, the matrix also allows us to approximate the mean pairwise correlations within the whole population at any given trial ($\check{R}_t$) by taking the mean over all unique neuronal pair values in matrix $\check{Z}$:

$$\check{R}_t = \Sigma_{i=[1 \ldots n-1]} \Sigma_{j=[i+1 \ldots n]} (\check{Z}_{i,j,t}) / ((n \cdot (n-1))/2) \tag{eq. 9}$$

Similarly, we can take the standard deviation instead of the mean over all unique pairs per trial to estimate the spread of the instantaneous pairwise correlation distribution. However, note that while the instantaneous Pearson-like correlation is similar to the conventional Pearson correlation, $\check{R}$ is not bounded within the interval [-1 1], because the z-scored element-wise product and the mean-operator work over different sets of values (i.e., matrix dimensions).

We additionally used for comparison a more conventional measure of correlations across time by using a wavelet-based sliding window correlation (Cooper and Cowan 2008). The time-scale of the wavelet used in all sliding-window analyses was set to 1.0 second, as this was similar to the animals' median reaction times and should therefore maximize the stimulus-driven change in neuronal pairwise correlations.

**Receiver Operating Characteristic (ROC) analysis of hit/miss separability**

We quantified the single-trial behavioral response predictability using an ROC approach by calculating the area under the curve (AUC) for a false positive rate vs. true positive rate plot. All ROC curves were computed separately per contrast and animal for both heterogeneity and mean population dF/F0 (fig. 3g). For comparison across animals, we averaged the AUC of the four test contrasts per animal, yielding a single AUC value per animal for both heterogeneity and dF/F0 (fig. 3h).

**Decoding of stimulus presence**

To ascertain the performance of a decoder on the same task as we required the mouse to perform, we created an algorithm that calculated the probability of a stimulus being present. This decoder was based on a previously published maximum-likelihood naive Bayes decoding algorithm (for a more complete description, see (Montijn, Vinck, and Pennartz 2014)). For each neuron and stimulus orientation, we computed the mean and standard deviation of mean dF/F0 during presentation of a 100% contrast stimulus as well as the mean and standard deviation during 0% probe trials. For each test trial and neuron with the preferred orientation as the trial's stimulus orientation, we calculated the probability a stimulus was present by reading out the likelihood density function for 0% and 100% contrast trials. The product over neurons in the preferred population for each trial then yields a population posterior probability value for stimulus absence (0% likelihood) and presence (100% likelihood). The decoder's read-out was the posterior with the highest probability. Because the likelihood was only based on 0% and 100% contrast responses, automatic cross-validation was ensured for decoding test contrast stimuli. After decoding stimulus presence for all trials, we split the trials into hits and misses and calculated the percentage for which the decoder indicated a stimulus was present per response type and contrast, averaging over repetitions and orientations. This yielded two curves per animal (see fig. 4d). We tested for statistically significant differences between response and no-response trials by performing a paired t-test over animals on the intermediate contrasts (0.5-32%).

Furthermore, we quantified the similarity of our decoder's performance to the animal's performance in the visual stimulus detection task, by calculating the similarity per animal of its actual behavioral performance to the decoder's performance (Pearson correlation over contrasts). We compared this value to the similarity obtained with a bootstrapped shuffling procedure (1000 iterations). Here, we shuffled the animal's behavioral and decoder performance over contrasts, recalculated the similarity index and took the mean over all iterations as the resultant shuffled similarity. To test for statistical significance, we performed a paired t-test over animals between the shuffled and real similarities (fig. 4e).

Moreover, we investigated the similarity between the animal's and decoder's output at a single-trial level with a chi-square analysis. Pooling all trials across animals showed significant correspondence between the decoder and animal's judgment of stimulus presence; hit trials were more often decoded as "stimulus present" and miss trials more often as "stimulus absent" (figure supplement 4-2j). Note that this decoding procedure is not optimal; the absolute decoding performance therefore should not be interpreted as reflecting the actual amount of information present in the neural responses. The purpose of this decoder is merely to test - in coarse terms - the similarity between the neural signal and the animal's behavior.

**Behavioral response predictability**

We analyzed the predictability of behavioral responses before they occurred based on either the mean population dF/F0 response or population heterogeneity between 3 and 0 seconds before stimulus onset (fig. 5e). Hit trials were split into the 50% fastest and 50% slowest reaction times per contrast per animal and then averaged over contrasts, yielding 6 data points per animal: the mean pre-stimulus population dF/F0 and mean population heterogeneity preceding fast, slow and miss trials. We then quantified the consistency of differences over animals by calculating the distance of these points per animal to the mean of their own response group and the other two. We defined the predictability metric per point *i* (animal) for two response types *r1* and *r2* (i.e., two types out of fast, slow or miss) as:

$$\delta_{r1,r2,i} = ((\|d(i_{r1},\mu_{r2})\|/(\|d(i_{r1},\mu_{r2})\| + \|d(i_{r1},\mu_{r1}^{\neg i})\|)) - 0.5) \cdot 2, \quad (eq.\ 10)$$

where $\|d\|$ is the absolute Euclidian distance (vector magnitude), $\mu_r$ is the mean location of $\mathbf{l}_r$ – where $\mathbf{l}_r$ is the group of points for response $r$ – and $\mu_r^{\neg i}$ indicates the mean location of $\mathbf{l}_r$ without point *i*. This analysis yields a vector $\boldsymbol{\delta}_{r1,r2}$; the separability between response type *r1* and *r2*. Random placement would lead to a separability of $\delta = 0$, so we quantified statistically significant predictability of responses by performing FDR-corrected one-sampled t-tests (vs. 0) for each separability vector and both neuronal population metrics (heterogeneity and mean dF/F0). We also tested whether the separability was higher for heterogeneity or dF/F0 by performing FDR-corrected paired t-tests between dF/F0 and heterogeneity separability vectors for the same response type comparisons (fig. 5e).

**Rise time to maximum heterogeneity**

We defined the rise time to maximum stimulus-driven heterogeneity as the time it took the population heterogeneity to rise from 10% to 90% of the difference between pre-stimulus baseline levels and maximum heterogeneity during the stimulus period. This rise time was calculated on the mean curves per animal and contrast as shown in fig. 5b. To create the graph shown in fig. 5f, we took the rise time across test contrasts per animal (n=8) and behavioral response type (miss, slow, fast). We tested for significant differences in average rise times between response types with paired t-tests across animals.

**Population activation pattern consistency**

Detection of a visual stimulus might be associated with consistencies in population activity. We therefore analyzed whether the inter-trial correlation of population activity varies depending on the behavioral performance of the animal. We again separated fast, slow and miss trials and for each stimulus orientation calculated the correlation of the dF/F0 response vector between pairs of trials with the same type of behavioral response (fig. 6a). We separated the neuronal responses for that orientation's preferred and non-preferred population of neurons, also to address whether consistency across trials might be restricted to the preferred population or would also occur in the non-preferred population (fig. 6b-d). Note that because we calculated the correlations separately for preferred and non-preferred populations, the relative contribution of the orientation signal is fairly low, which explains the relatively low correlation values. To assess above-chance similarities, we compared these values to correlations obtained from shuffled data. By shuffling within each stimulus orientation all trial identities randomly for each neuron, the orientation signal is preserved, but other similarities across trials are destroyed. We repeated this shuffling procedure 100 iterations and took the mean of these 100 iterations as shuffled correlation value per animal (fig. 6b-d). To test for statistically significant consistencies in population activation patterns, we performed FDR-corrected paired t-tests between the real and shuffled correlation values over animals for the different response types and the two neuronal population types. We also quantified the differences between response groups in the real data with paired t-tests (miss vs. slow, miss vs. fast, and fast vs. slow).

**Analysis of multidimensional inter-trial distance in neural activity**

To study the theoretical implications of our results relating to heterogeneity, we proceeded with an analysis of the question whether heterogeneity forms a special case of population codes that do not merely reflect an increased activity of all neurons upon visual detection. For the specific purpose of these analyses (shown in fig. 7), we use as definition for multidimensional heterogeneity the distance in neural space from the population's activity to the closest point on the main diagonal (see text and below for further explanation). Although this definition is computationally different our pairwise definition of heterogeneity, it also captures the overall dissimilarity of responses within a population of neurons. Moreover, applying this procedure to z-scored dF/F0 values yields Pearson's correlations of r > 0.9 when compared with our original definition of heterogeneity (eq. 3-4), and gives very similar hit/miss Cohen's d values (figure supplement 3-2). The two metrics therefore likely capture the same neural phenomenon and show that heterogeneity can be studied by different, but related computational definitions.

To assess the distribution of neuronal population activity in multidimensional neural response space (where each dimension represents the activity of a single neuron (see fig. 7a-d)), we calculated the inter-point distance (each point representing the population activity during a single trial) between all hit trial pairs and between all miss trial pairs. The distance in neuronal activity for a population of *n* neurons between a pair of trials *x* and *y* in multidimensional space can be calculated as the *n*-dimensional Euclidian:

$$\boldsymbol{d(x,y)} = \sqrt{(x_1 - y_1)^2 + (x_2 - y_2)^2 + \ldots + (x_n - y_n)^2} \qquad \text{(eq. 11)}$$

The pairwise inter-point distance is then given in units of neural activity (dF/F0, fig. 7e). Note that this formula can also be used to calculate the multidimensional heterogeneity, as defined above, by taking the distance between any trial (*x*) and the closest point on the diagonal (*y*).

Next we investigated the symmetry of population responses around the main diagonal, as this symmetry gives an indication of whether heterogeneity is an epiphenomenal observation, or a fundamental neural characteristic underlying visual detection (see text). In order to do so, we mirrored each point across the diagonal and recalculated the inter-point distances for the mirrored data. Mirroring across the diagonal was achieved by direct inversion of the signs per neuron relative to the main diagonal. For a population response $r=[r_1\ r_2\ ...\ r_i\ ...\ r_n]$, where n is the number of neurons, the mirrored version $r'=[r'_1\ r'_2\ ...\ r'_i\ ...\ r'_n]$ was calculated as follows:

$$r' = (\mu_r - (r - \mu_r)) \qquad (\text{eq. 12})$$

where $\mu_r$ is the mean population response over $r$.

**Removal of mean and heterogeneity, and subsequent hit/miss decoding**

For the analyses displayed in fig. 7g,h, we removed the mean and/or heterogeneity from the population responses and assessed the effect on decoding accuracy of hit/miss responses during test contrast stimuli. As mentioned before, for these analyses heterogeneity was defined as the distance to the main diagonal. As such, removal of the mean without influencing heterogeneity is trivial and can be achieved by simply subtracting the mean population response from all neuronal dF/F0 values obtained for each trial. Briefly, heterogeneity was removed from each trial without affecting the mean in two steps; first heterogeneity was removed, and next any influence on the mean was remedied by adding the difference between the new and old mean. First, heterogeneity was removed by dividing each neuron's response during that trial by the square root of the sum of the squared differences between the neuronal responses and the mean (i.e., by dividing by the heterogeneity):

$$r'^{\neg H} = \frac{r}{\sqrt{(r_1-\mu_r)^2 + (r_2-\mu_r)^2 + ... + (r_n-\mu_r)^2}} \qquad (\text{eq. 13})$$

Next, changes in the mean were corrected by removing the new mean of the heterogeneity-removed population activation ($\mu_{r'^{\neg H}}$) and adding the old population mean $\mu_r$:

$$r^{\neg H} = r'^{\neg H} + \mu_r - \mu_{r'^{\neg H}} \qquad (\text{eq. 14})$$

This way, the heterogeneity (i.e., the Euclidian distance of that trial's population activity to the main diagonal) is normalized to 1.0 for all trials. The multidimensional location relative to the diagonal is preserved, but its distance is always the same; all trials now fall on a cylinder with a radius of 1.0 dF/F0 around the main diagonal. In other words, the population activation during a trial is projected as a vector from the closest point on the diagonal to the trial's position, and the vector's angle is preserved, but its magnitude is normalized to 1.0. Both properties (mean and heterogeneity) can be removed by subtracting the mean from the heterogeneity-removed responses. Removing the mean as well as the heterogeneity collapses this cylinder onto a circle through multidimensional space around the origin.

**Control analyses for confounds related to running, licking and eye movements**

To control for potential locomotor confounds, we split all datasets into trials where the mouse was still (90.9 +/- 3.6% of trials) and where it was moving during stimulus presentation (8.1% +/- 3.6% of trials), and reanalyzed our data. Our results with exclusion of running trials (figure supplement 2-1g,h) are very similar to our original analysis (fig. 2a,b), showing that the effects we observed cannot have been due to running-induced modulations (paired t-test, hit vs miss, 0.5% - 32%, p<0.05).

Another potential confound for our results could be that response trials induce signals related to motor feedback or motivation to initiate motor actions, because the animal initiates licking as a behavioral response. This also seems unlikely; because 0% contrast probe trials did not induce neuronal activity during false alarms (fig. 2a, figure supplement 2-1a, green line). Theoretically however, such signals could still be present and influence population activity only when occurring concurrently with visual stimulation. To control for this, we re-performed our analyses shown in fig. 2a,b, but now used data only from the first 0.4s after stimulus onset;

approximately 0.8s before the mean reaction time. Leaving a window of 0.8s between the latest frame included in the data analysis and the licking response should also eliminate potential modulatory activity from motor cortex related to the preparation of licking. The results from this control analysis were slightly noisier, due to the shorter data acquisition duration per trial, but showed no qualitative differences to the original analysis regarding heterogeneity (figure supplement 2-1i,j). The intermediate contrasts still showed significant enhancements in heterogeneity ($p<0.01$) during hit trials, but we found no significant differences for mean population dF/F0 ($p=0.543$). We therefore conclude that our results regarding heterogeneity are not confounded by motor-related modulations due to running or licking, nor by reward-expectation prior to licking responses, and confirm that the mean population dF/F0 is not or less useful as a measure of neural correlates of perception.

To control for possible effects of blinking and saccades, we performed pupil detection on our eye-tracking data and removed all trials in which the animals blinked or made saccades during any time of the stimulus presentation (10.2% +/- 4.6% of trials removed (mean +/- standard deviation)). We re-performed our analyses on only the trials where no contamination by incorrect eye position and/or closing of the eyelids was possible (figure supplement 2-1k,l) and observed that our results regarding heterogeneity were qualitatively and quantitatively similar to our original analyses, but that the dF/F0 results were again more sensitive to a conservative analysis (hit/miss difference for intermediate contrasts, paired t-test, n=8; dF/F0, $p=0.136$; heterogeneity, $p<0.005$). We conclude that our main results are not biased by incorrect eye position and blinking.

**Orientation decoding**

We addressed whether the orientation information contained in the population responses was dependent on the mean dF/F0 and heterogeneity during stimulus presentation. We decoded the presented stimulus orientation for each contrast separately (i.e., 100% contrast trials based on likelihood from 100% contrast trials, etc.) by a leave-one-out cross-validation and afterwards split all trials into correctly and incorrectly decoded ones (figure supplement 4-2b). To quantify the dependence of decoding accuracy on dF/F0 during stimulus presentation, we took for each contrast the trials with highest and lowest 50% of dF/F0 and calculated the mean decoding accuracy for both groups (high and low activity). Next, we took the mean for these groups over contrasts per animal and calculated a percentage decoding accuracy increase for the highest vs. lowest 50% dF/F0 trials (see figure supplement 4-2c). To test for statistical significance, we performed a one-sample t-test of the percentage increase values over animals. For heterogeneity, we performed the same steps and performed a t-test vs. 0% increase (figure supplement 4-2c).

**Stimulus feature decoding**

To address whether visual stimulus features (i.e., orientation and contrast) were more accurately represented by neuronal population activity during correct vs. incorrect behavioral performance, we used a Bayesian maximum-likelihood decoder as previously described to extract those features from the population activity (for a more complete description, see (Montijn, Vinck, and Pennartz 2014)). We defined all combinations of orientations and contrasts as different stimulus types, yielding a total of 21 different stimulus types (four orientations times five contrasts plus probe trials). Next, we performed a leave-one-out cross-validated decoding procedure for all trials and calculated the mean percentage correct decoding trials for hits and misses per stimulus type; then we averaged the percentage correct over stimulus types, yielding an accuracy per animal for hit and miss trials. We tested for a statistical difference between hits and misses with a paired t-test over animals (fig. figure supplement 4-2a).

**Noise correlations**

To investigate detection-related increases or decreases in noise correlations (figure supplement 4-2f,g), we first calculated a response vector for each stimulus orientation θ that was presented during a test contrast trial. Here, each element in the vector is the neuron's response to a single presentation *t* (i.e., a trial) of that stimulus orientation:

$$\boldsymbol{R}_\theta = [R_{\theta\,t} \;...\; R_{\theta\,n}] \qquad\qquad\qquad\qquad\qquad\text{(eq. 15)}$$

where *n* is the number of repetitions per response type per orientation. Because we aim to compare a single noise correlation value per neuronal pair *i,j*, we took the mean noise correlation over all four stimulus orientations:

$$\rho_{i,j}^{\text{noise}} = \frac{\sum_{\theta=0}^{135} \text{corr}(\boldsymbol{R}_{i,\theta},\boldsymbol{R}_{j,\theta})}{4} \qquad\qquad\qquad\text{(eq. 16)}$$

The noise correlation is therefore an index of the mean trial-by-trial variability shared by pairs of neurons over all stimulus orientations.

**Behavioral response predictability on single trial basis**

To verify that the behavioral predictability before stimulus onset that we found (fig. 5e) was not merely a group-level effect, but was indeed also a single-trial phenomenon, we subsequently performed single-trial decoder-based predictions of fast/slow/miss behavioral responses that occurred during the subsequent stimulus presentation (see figure supplement 5-1). We used a similar leave-one-out cross-validated naive Bayes decoder as described above for fast, slow and miss trials, and calculated per trial the relative likelihood that the subsequent stimulus presentation would lead to a miss, fast or slow response. We then split the predictive decoding results per actual behavioral response group and averaged the relative prediction likelihood per animal. This yields 3 relative probability values per actual response type per animal. Assigning an angle to each of these behavioral responses that are separated by $2/3\pi$ on the unit circle and taking the relative likelihood as the vector magnitude, it is then possible to calculate a resultant prediction vector per actual response type per animal. To quantify statistical significance, we multiplied an angle-based correctness index (+1 when the resultant prediction vector angle is perfectly aligned to the actual response angle and -1 when they are separated by $1\pi$) with the vector magnitude, giving a normalized decoding accuracy index between -1.0 and +1.0, where chance level is 0. Lastly, we performed one-sample t-tests on the normalized decoding accuracy indices over animals and response types for heterogeneity and dF/F0, and a paired t-test between dF/F0 and heterogeneity (figure supplement 5-1).

**Author Contributions** J.S.M. and P.M.G. built the setup. J.S.M. performed the experiments and analyzed the data. J.S.M. and C.M.A.P. designed the experiments and analyses, and wrote the paper.

**Acknowledgements** We thank Q. Perrenoud, G. Meijer and M. Vinck for feedback on earlier versions of the manuscript and fruitful discussions. We would also like to thank W. Oldenhof, J. Verharen and L. Forsman for assistance with training the animals. This work was supported by the Netherlands Organization for Scientific Research-Excellence grant for the Brain & Cognition project 433-09-208 and the EU FP7-ICT grant 270108. The authors declare no competing financial interests.

**Figure legends**

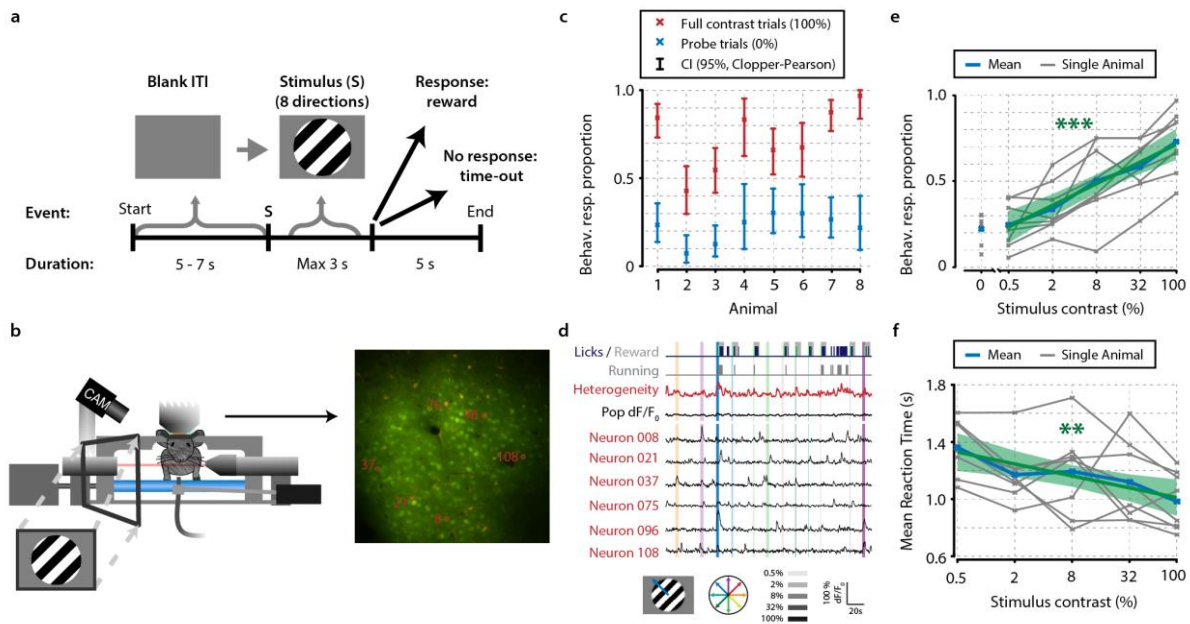

Figure 1. Mice perform a go/no-go task during *in vivo* calcium imaging. **a**, Task schematic showing the time course of a single trial. In each trial, one of a combination of eight different directions and five contrasts, or a 0% contrast probe trial (isoluminant grey blank screen), was presented (ratio 1:5 of 0%-contrast-probe:stimulus trials). When mice made a licking response during stimulus presentation the visual stimulus was turned off and sugar water was presented. **b**, Schematic of experimental setup. During task performance we recorded eye movements with an infrared sensitive camera, licking responses and running on a treadmill. **c**, All eight animals performed statistically significant stimulus detection during neural recordings, as quantified by non-overlapping $2.5^{th}$ - $97.5^{th}$ Clopper-Pearson percentile confidence intervals (95% CI) (p<0.05) of behavioral response proportions for 0% and 100% contrast probe trials. **d**, Example of simultaneously recorded behavioral measures, population heterogeneity, mean population dF/F0, and traces of neurons labelled in panel **b**. Vertical colored bars represent stimulus presentations; width, color and saturation represent duration, orientation, and contrast respectively. **e,f**, Animals showed significant increases in behavioral response (behav. resp.) proportion (linear regression analysis, see Materials and Methods, p<0.001) (**e**) and reductions

in reaction time (p<0.01; **f**) with higher stimulus contrasts. Shaded areas show the standard error of the mean. Statistical significance; ** p<0.01; *** p<0.001.

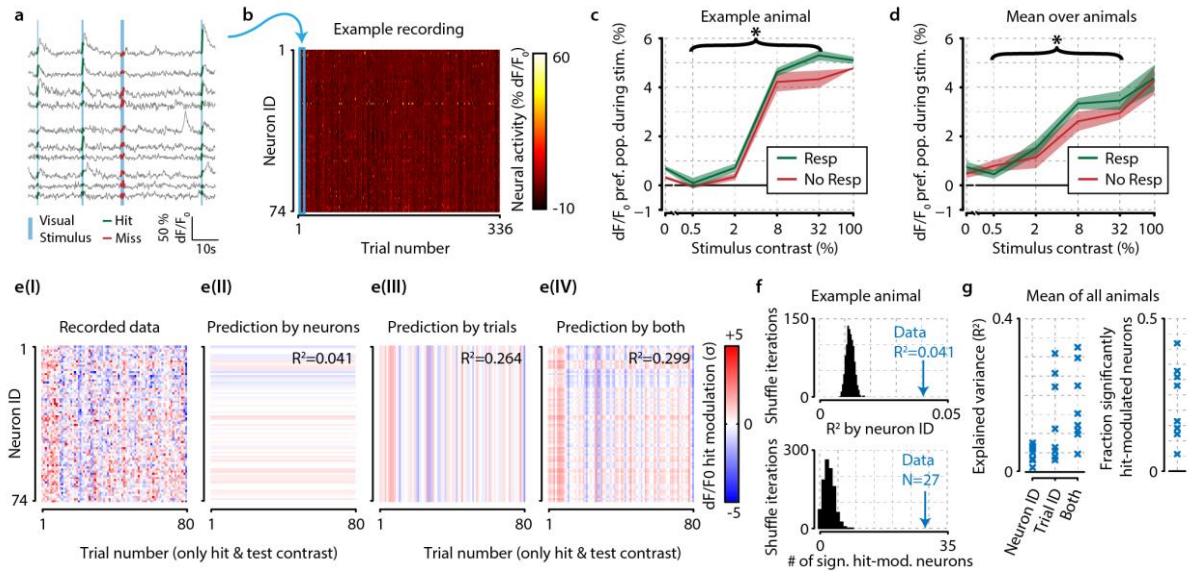

Figure 2. The difference in neural activity between hit and miss trials can be partly explained by consistent hit-associated increases in activity of specific neurons, and somewhat better by trial-by-trial population-wide fluctuations, but these mean-based approaches fall short of being fully descriptive. **a**, Recorded traces from 10 randomly selected neurons over four subsequent trials. For further analysis, we took the mean dF/F0 per neuron over the visual stimulation period (thick colored lines) as single mean neural activity measure per trial. **b**, Data of one entire recording block consisting of 74 tuned and simultaneously recorded neurons over 336 trials. Blue rectangle shows the four trials depicted in panel a. **c**, In an example animal, the detection of stimuli (green) with test contrasts (0.5 – 32%) correlated with a modest increase in preferred population dF/F$_0$ over undetected stimuli (red) (two-sample t-test, p<0.05), but none of the individual contrasts reached statistical significance (resp. vs. no resp., two-sample t-tests, FDR-corrected p>0.05). **d**, As c, but for mean over all animals the graph shows a small, but consistent overall difference of dF/F$_0$ with visual detection (test contrasts 0.5% - 32%, n=8 animals, p<0.05). **c,d**, Shaded areas show the standard error of the mean. **e**, The hit-associated increase in neural activity per neuron for all hit trials of test contrast stimuli (panel I) can be partly explained by specific neurons showing consistent dF/F0 increases or decreases across trials (panel II), partly by trial-by-trial population-wide fluctuations regardless of neuronal

identity (panel III), and somewhat better by both (panel IV). **f**, Control analysis by shuffling neuronal identities (IDs) per trial (n=1000 iterations, black distribution) shows that the population activity is more predictable based on consistent hit-modulations per neuron (top panel) and more neurons are significantly hit-modulated (bottom panel) than can be expected by chance. **g**, Analyses as in f, but across animals; comparison vs. shuffle-based $R^2$-expectation showed above-chance (at α=0.05) predictability of hit-modulations using neuron ID, trial ID or both for respectively 7/8, 8/8 and 8/8 animals (left panel). The fraction of significantly hit-modulated neurons was above-chance (at α=0.05) for 7/8 animals (right panel).

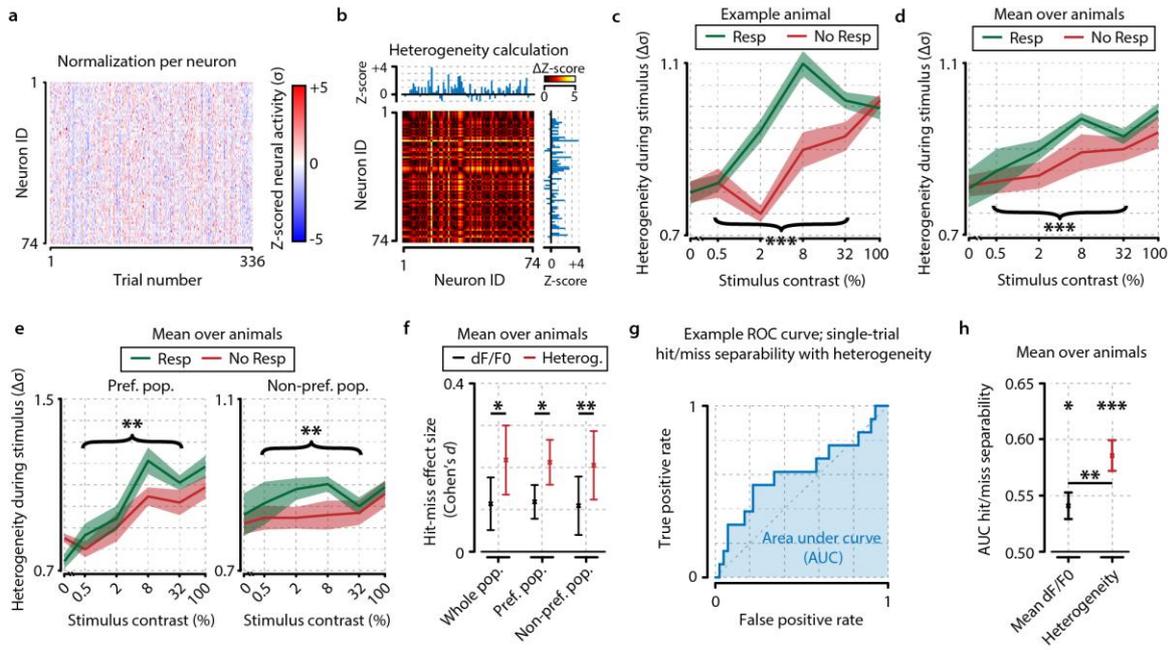

Figure 3. Neuronal response heterogeneity within populations correlates better with visual detection than mean preferred population (pref. pop.) activity. **a**, Neuronal activity as in fig. 2b, but presented as z-score normalized per contrast to be able to compare relative changes across contrasts. **b**, Schematic representation of the method to compute heterogeneity on an example trial (see also Materials and Methods). The dF/F0 response of each neuron is z-scored per contrast and the distance (absolute difference) in z-scored activity between all pairs of neurons is calculated for each trial (color-coded ΔZ-score). The population heterogeneity in a given trial is defined as the mean ΔZ-score over all neuronal pairs. **c**, Population activity heterogeneity in an example animal shows a strong correlation with visual detection. Comparison between detected (resp.) and undetected (no resp.) trials for test contrasts as a group (paired t-test, p<0.001) was highly significant. **d**, As c, but showing mean over all animals (n=8). Stimulus detection correlated with higher heterogeneity; test contrast group hit-miss comparison was highly significant (p<0.001). **e**, as d, but for heterogeneity within the preferred (left panel) and within the non-preferred (right panel) population only. Hit-miss differences were found in the preferred population (test contrast group, p<0.01) and non-preferred population (test contrast group, p<0.01), similar to the whole population (panel d). **f**, Using a measure of effect size

analysis (Cohen's *d*), heterogeneity was found to show a stronger correlation with stimulus detection than mean dF/F0 within the whole population (Cohen's *d*=0.114 vs d=0.218*)*; within the preferred population (Cohen's *d*=0.119 vs d=0.213) and within the non-preferred population (Cohen's *d*=0.110 vs d=0.206*)* (paired t-tests over animals (n=8), whole population; p<0.05, preferred population; p<0.05, non-preferred population; p<0.01). **g**, Example Receiver Operating Characteristic (ROC) curve showing the linear separability of single-trial hit and miss trials using population heterogeneity (see Materials and Methods). The separability can be quantified by the area under the curve (AUC; blue shaded area). True positive rate: fraction of hit trials classified as hit. False positive rate: fraction of miss trials classified as hit. **h**, Statistical quantification of hit/miss separability using either mean dF/F0 (black) or heterogeneity (red) across animals (n=8). Both measures predict the animal's response above chance (FDR-corrected paired t-test dF/F0 and heterogeneity AUC vs. 0.5, p<0.05 and p<0.001 respectively), but behavior can be predicted better using heterogeneity (paired t-test, dF/F0 vs. heterogeneity AUC, p<0.01). All panels: Shaded areas/error bars show the standard error of the mean. Statistical significance; * p<0.05; ** p<0.01; *** p<0.001.

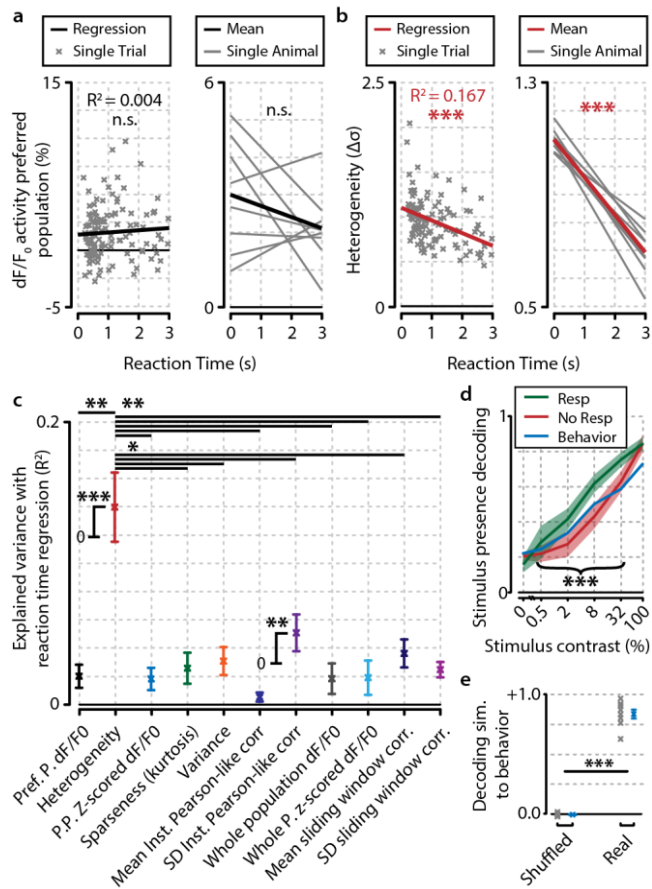

Figure 4. Heterogeneity is correlated with reaction time. **a**, Reaction time shows no correlation with mean preferred population dF/F0 activity during stimulus presentation for any individual animal (left panel; example animal), nor for the meta-analysis over animals (right panel; FDR-corrected one sample t-test over individual regression slopes per animal, n=8, n.s.). **b**, Same as **a**, but for heterogeneity. Heterogeneity shows a strong correlation with reaction time (left panel; example animal, p<0.001) as well as well for the meta-analysis over animals (p<0.001). **a,b**, Note different y-axis scaling per panel for display purposes. **c**, Comparison of the explained variance of several neural metrics. Only heterogeneity (FDR-corrected one sample t-test, p<0.001) and spread (SD; standard deviation) in instantaneous Pearson-like correlation (see Materials and Methods) (p<0.01) correlate significantly with reaction time; all other metrics do not (preferred-population (Pref. P.) dF/F0, preferred-population (P.P.) z-scored dF/F0, variance, sparseness (kurtosis), mean instantaneous Pearson-like correlation, whole-population

(z-scored) dF/F0, mean and SD of sliding window correlation (width 1.0s); all n.s.). Heterogeneity explains more reaction-time dependent variance than any other metric (FDR-corrected paired t-tests, all $p<0.05$). **d**, Decoding of stimulus presence shows similar accuracy as actual behavioral performance by the animals (fig. 1e). When the animal has detected the stimulus (resp; green line), the decoder is better able to correctly judge its presence (a value of 1 indicates perfect performance, paired t-test, $p<0.001$). Shaded areas show the standard error of the mean. **e**, Behavioral detection performance is more similar (sim.) to the optimal decoder's performance than expected by chance (paired t-test, n=8 animals, shuffled vs. real similarity, $p<0.001$). Grey: single animal; blue: mean across animals. All panels: error bars/shaded regions show standard error of the mean. Statistical significance; * $p<0.05$; *** $p<0.01$; *** $p<0.001$.

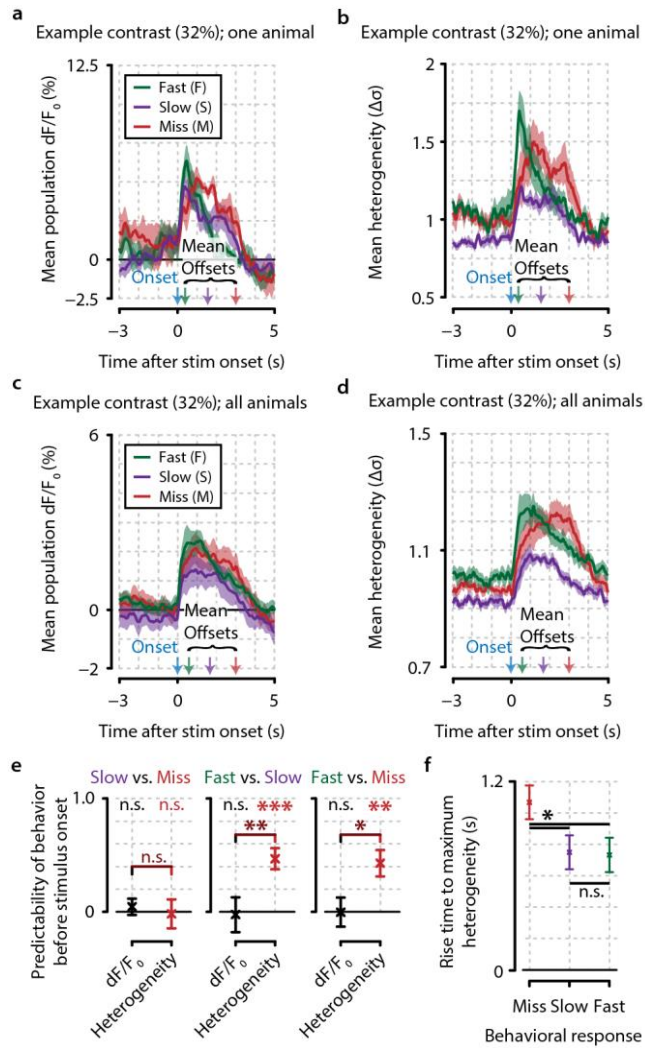

Figure 5. Heterogeneity preceding stimulus onset predicts behavioral reaction time. **a**,**b**, Example traces from one example animal for population dF/F0 (a) and heterogeneity (b) of fast (F, green), slow (S, purple) and miss (M, red) responses (mean +/- standard error over trials) also showing mean stimulus offsets. **c**,**d**, As **a**,**b**, but showing mean +/- standard error over animals. **e**, Fast behavioral responses are predictable before stimulus onset using heterogeneity (FDR-corrected one-sample t-tests vs chance level (0); S-M, p=0.799; F-S, p<0.001; F-M, p<0.01), but not using dF/F0 (FDR-corrected one-sample t-tests vs chance level; S-M, p=0.157; F-S, p=0.811; F-M, p=0.924; FDR-corrected paired t-test for heterogeneity vs dF/F0; S-M, p=0.477; S-F, p<0.01; F-M, p<0.05). **f**, The population heterogeneity during stimulus presentation does not merely reflect a continuation of pre-stimulus neural state; detected stimuli

(slow and fast responses) elicit a faster rise to the maximum heterogeneity level than undetected stimuli (miss trials) (paired t-tests, n=8 animals, p<0.05). Slow and fast responses do not differ significantly (p>0.05). All panels: Error bars/shaded areas indicate standard error of the mean. Statistical significance; * p<0.05; ** p<0.01; *** p<0.001.

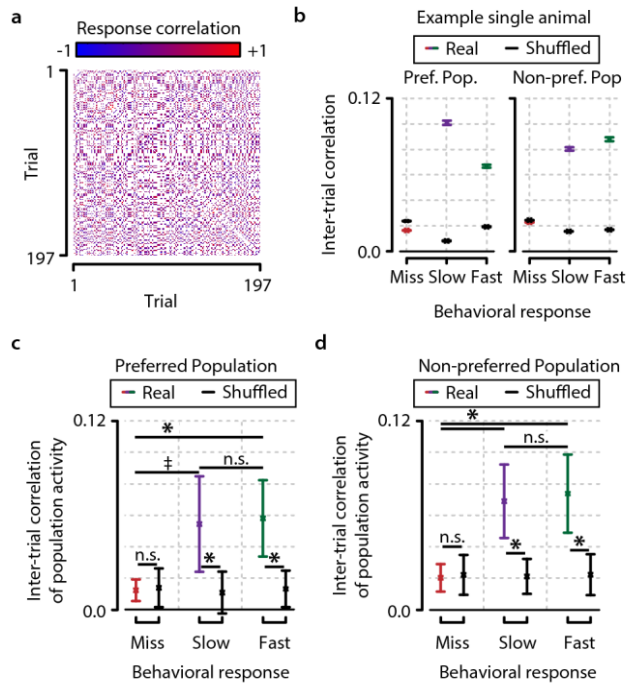

Figure 6. Consistency in population activation patterns across trials is increased during hit trials (fast and slow) compared to miss trials. **a**, Data from example animal showing inter-trial correlations (Pearson's r) between population responses to same-orientation stimuli (pooled over test contrasts only). **b**, Data from same animal as in panel a, showing population higher activity pattern consistency (mean +/- standard error over trial pairs) for fast and slow response trials than miss trials within both the preferred and non-preferred population. Colored lines show real data, black lines show shuffled data (see text and Materials and Methods). **c,d**, Inter-trial correlations (mean +/- standard error over all animals) as quantification of population activation pattern consistency are significantly higher during fast trials (with a trend for slow trials) than during miss trials within the preferred as well as the non-preferred neuronal population, suggesting that visual stimulus detection is correlated with the occurrence of more stereotyped population responses (FDR-corrected paired t-tests, preferred population; miss-slow, p=0.081; miss-fast, p<0.05; slow-fast, n.s.; non-preferred population; miss-slow, p<0.05; miss-fast, p<0.05; slow-fast, n.s.). Comparison with correlations of shuffled data yielded similar results (both preferred and non-preferred populations; paired t-test real vs. shuffled: miss;

p>0.2, slow and fast; p<0.05). Error bars indicate standard error. Statistical significance; * p<0.05; ǂ 0.05<p<0.1.

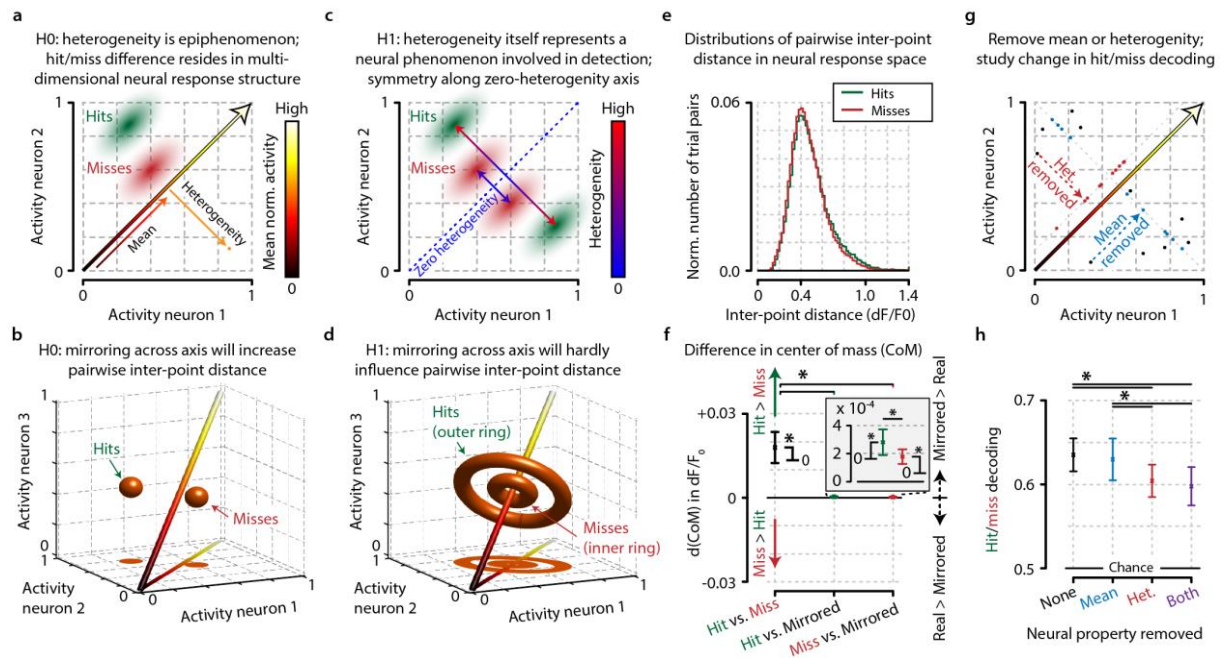

Figure 7. Conceptual interpretation of heterogeneity as neuronal population coding phenomenon. **a-d**, The mean population activity during a certain time epoch can be visualized a single point in multidimensional neural space, where every axis represents the activity of a single neuron. **a**, For an example population of two neurons, the main diagonal (arrow) represents the line along which the mean population activity changes. Orthogonal to this line is the gradient along which heterogeneity changes, representing the distance of each point to the main diagonal. The effects of heterogeneity on hit/miss differentiation as reported in this study could be epiphenomenal if the real underlying differentiation depends on localized, segregated clusters of neural activity for hits (green cloud) and misses (red cloud). **b**, This principle can be extended to multidimensional space; segregated clusters activity will show asymmetrical distributions of population activity around the diagonal. **c,d** Alternatively, heterogeneity itself could represent a fundamental characteristic of hit/miss differences; in this case population responses should be distributed symmetrically around the diagonal (see text for more explanation). **e**, Calculating the pairwise inter-point distance (each point being the population activity during a single trial) can reveal information about the underlying multidimensional structure of neuronal population activity. Green: distribution of inter-point distances for hit

trials; red: same for miss trials. **f**, Population responses during hit trials are distributed within a larger volume of neural space, as shown by the on average larger inter-point distance for hits than misses (paired t-test, difference in center of mass (d(CoM)) hit vs. miss, $p<0.05$). Mirroring point across the diagonal to assess symmetry shows a small, but significant asymmetry for hits and miss (both $p<0.05$), and larger asymmetry for hits than misses ($p<0.05$). This suggests that neuronal populations during hit trials show more structured behavior in a more extended neural space than during miss trials. **g**, Schematic representation of how the mean, heterogeneity, or both can be removed from population responses to assess the effect they have on hit/miss separability (see also text and Materials and Methods). **h**, Removing heterogeneity impairs hit/miss decoding more than removing the mean (paired t-test, $p<0.05$), but in all cases (including removing both) the hit/miss separability is still well above chance (0.5). This suggests that heterogeneity is more important than population mean activity for differentiating stimulus detection from non-detection, but that other more complex neural phenomena account for most of the population response structure. All panels: error bars indicate standard error. Statistical significance; * $p<0.05$.

**Supplementary figure and video legends**

Video 1. Typical raw data example showing the first 10000 recorded frames from animal 5. The left hand side shows xy-corrected, but otherwise unaltered, raw fluorescence data. The legend above this raw data video shows the time after start of the experiment in seconds and the number of acquired frames. The top two panels on the right hand side show (left) a depiction of the stimulation screen (grey isoluminant background or oriented drifting grating during stimulus presentation), and (right) whether the mouse is making a licking response. The two panels below show a live updated summary of mean dF/F0 (left) and heterogeneity (right) during each trial. Green indicates a licking response, and red indicates no response. The two lower panels show a live trace of mean population dF/F0 and heterogeneity. Licking responses are shown as red dotted lines, and stimulus presentations are shown as a grey shaded area. Note that the recording is very stable, except during periods of heavy licking, such as after hit responses, when reward is delivered. Also note that neural data acquired during licking is not used for any of our analyses and does therefore not influence our results. The mouse is licking vigorously during the initial period of the recording, but more typical behavior sets in less than two minutes after start of the recording. Near the end of the video, it can be seen that hits and misses are more easily separable using heterogeneity than dF/F0 (although this difference is stronger in the example video than in the entire data set as a whole).

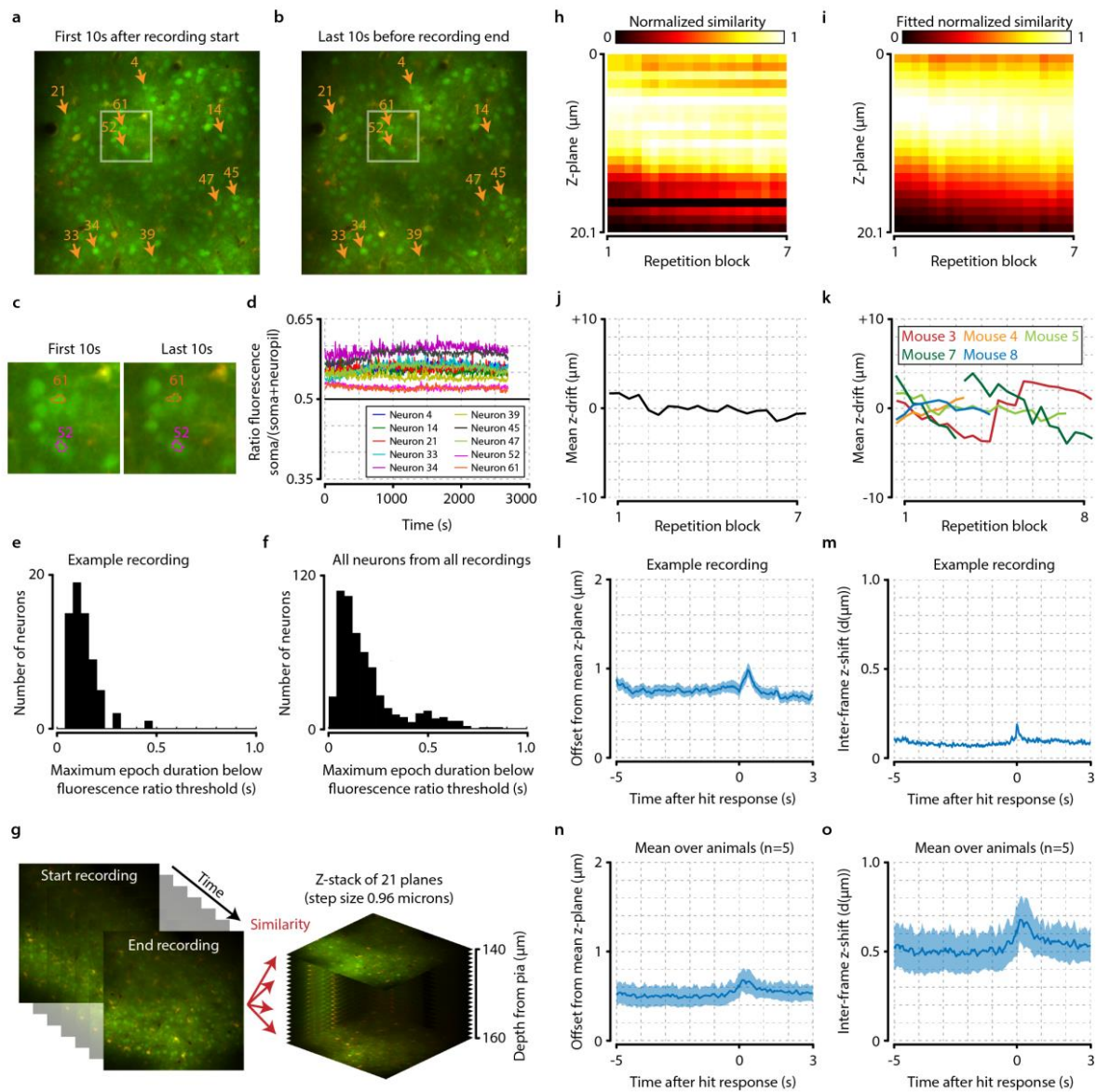

Figure supplement 1-1. Neuronal signals are stable over time. **a,b**, example from one animal showing the average of all recorded frames during the first 10 seconds after imaging onset (**a**) and during the last 10 seconds before the end of the recording (**b**). The locations of 10 randomly selected neurons are marked by orange arrows, and the white colored rectangles depict the area enlarged in panel **c**. **c**, enlarged subsection of the images shown in **a** and **b**, showing the outlines of cells 52 and 61. Note the similarity of the cell bodies and their outlines. **d**, down-sampled traces showing the ratio of soma vs. neuropil fluorescence of the 10 randomly selected neurons shown in panels **a** and **b** over duration of the entire recording (see Materials and Methods). All traces remain above the equiluminance threshold of 0.5, indicating that these neurons' somata remain visible and do not disappear due to bleaching or slow z-drift. **e**, histogram showing the maximum epoch duration of dropping below the equiluminance threshold of 0.5 for all neurons from the example session of a-d. None of the cells remain undetectable for longer than 0.5 seconds. **f**, as **e**, but for all cells from all recordings. None of the cells have a prolonged period (>1.0 second) of being undetectable, indicating stability of the neuronal signals over time. **g**, Schematic showing procedure of z-drift calculation for example animal. For each repetition block of 48 trials, we compared the similarity of 100 frames (~4 seconds) in the beginning, middle and end of each stimulus repetition set to slices recorded at different cortical depths (step size 0.96 microns). **h**, Similarity values normalized between [0-1] for each time point show a small and slow z-drift over the entirety of the recording. **i**, To quantify this drift, we fitted each time point with a Gaussian. **j**, The mean of the fitted Gaussian shows that maximum drift in z-direction for this animal did not exceed four microns. **k**, Data for all animals of which we recorded z-stacks (n=5/8). Animals 2 (not shown) and 7 showed excessive z-drift, so we split their recordings into two populations (see Materials and Methods). Within a single recorded population, z-drift never exceeded 8 microns for any animal. **l-o**, Frame-by-frame analysis of fast z-shifts during entire recordings show that z movement rarely exceeds more than a couple of microns. **l**, Example recording showing mean +/- standard error across trials in offset from mean z-plane as a function of time after hit response (licking). Shortly after the behavioral response, a peak in variability can be observed. However, note that for all analyses we only included epochs before the licking response; the period of neural responses with potential contamination from z-shifts will therefore not influence our results. **m**, Example frame-by-frame shift in microns (mean +/- standard error across trials) shows a similar time course as the mean z-offset. **n,o**, Same as panels l,m, but showing mean +/- standard error across animals (n=5).

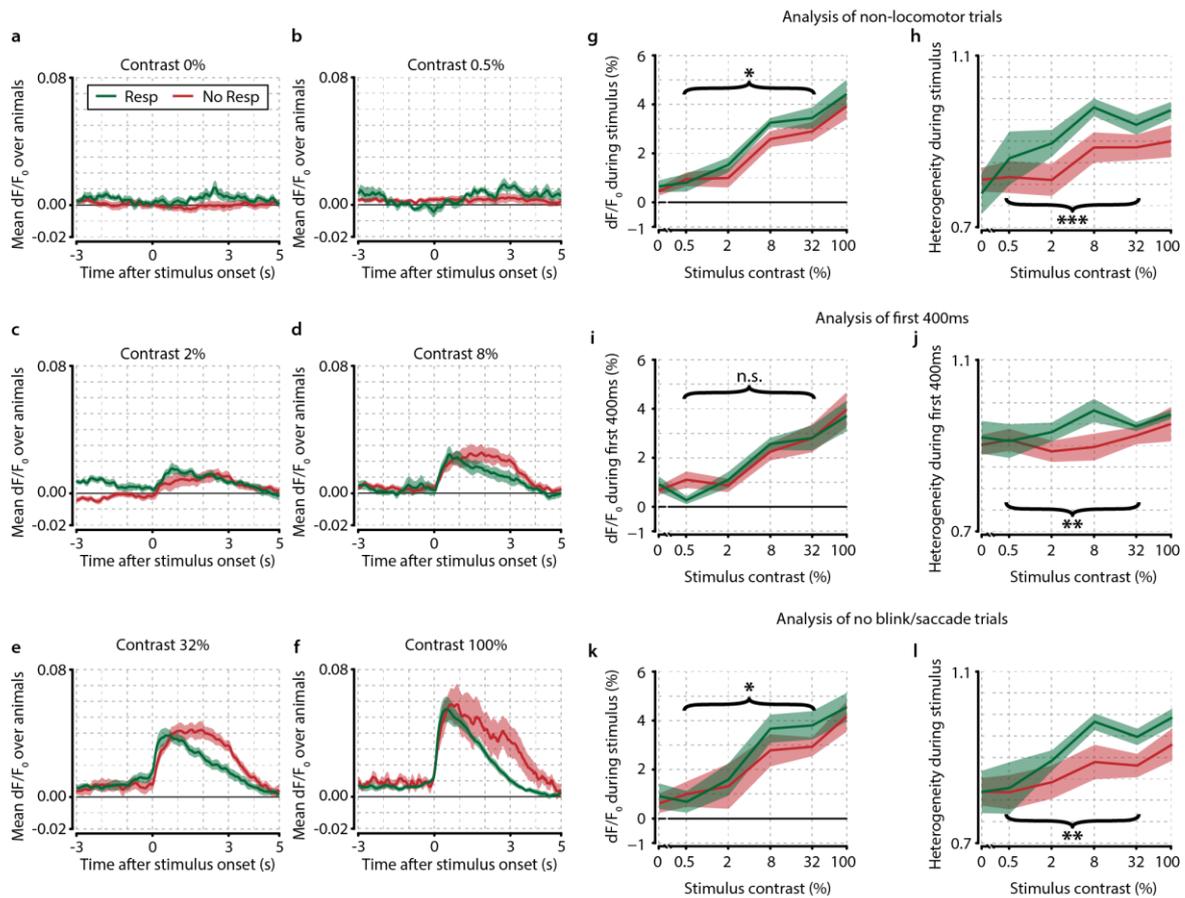

Figure supplement 2-1. Several control analyses reveal no confounding effects of motor preparation, modulatory feedback and eye movements. **a-f**, Traces of population dF/F0 averaged over animals (n=8) for different contrasts show the difference between response and no-response trials over time. **a**, probe trials (0% stimulus contrast). Note the absence of any neuronal activity during responses to 0% contrast probe trials (green line, t-test vs. 0, p=0.549) **b**, 0.5% contrast. **c**, 2% contrast. **d**, 8% contrast. **e**, 32% contrast. **f**, 100% contrast. **a-f**, Response trials (green) show quicker offsets, because the stimulus turns off when the animal makes a licking response. Also note the absence of any motor-related neural activity in panel **a**, supporting the interpretation that the observed correlates are unrelated to motor activity, preparation or reward expectancy. **g,h**, Removal of locomotion trials does not qualitatively affect neural correlates of stimulus detection. **g**, When computed only on trials where animals were not moving (89.9% of trials), the mean population dF/F0 as a function of stimulus contrast

shows little difference with the original analysis (compare to fig. 2d). Paired t-test over test contrasts (0.5%-32%) showed a significant difference between response and no-response trials ($p<0.05$). Our original results are very similar to the current analyses and are therefore not dependent on movement-induced modulations. **h**, As **g**, but for heterogeneity (still trials only); the overall paired t-test for hit/miss differences grouping 0.5% - 32% contrasts was highly significant ($p<0.001$), suggesting that our main results are not due to locomotion-related artefacts. **i,j**, Population correlates of visual detection are not dependent on motor-related and/or feedback signals. **i**, As fig. 2d; mean population dF/F0 as a function of contrast, now using only the first ~400 ms (394 ms; 10 frames) after stimulus onset. Mean reaction time over animals and contrasts was ~1.2s (see fig. 1f), leaving on average about 0.8s between the last data point included in this analysis and the subsequent licking response. **j**, As **i**, but for heterogeneity (compare with fig. 3d). Results were qualitatively similar to our original analysis for heterogeneity, but not for dF/F0 (paired t-test over intermediate contrasts for dF/F0, $p=0.543$, for heterogeneity, $p<0.005$). **k,l**, Population correlates are not explained by eye blinks or saccades. **k**, As fig. 2d; mean population dF/F0 as a function of contrast, using only trials in which the animal's eye position remained fixed and no blinks were detected during the entire stimulus period. **l**, As **k**, but for heterogeneity (compare with fig. 3d). Our results are qualitatively and quantitatively similar for dF/F0 and heterogeneity (hit-miss paired t-test over test contrasts; $p<0.05$ and $p<0.005$ respectively). All panels: Shaded areas show the standard error of the mean. Asterisks indicate statistical significance; * $p<0.05$; ** $p<0.005$; *** $p<0.001$.

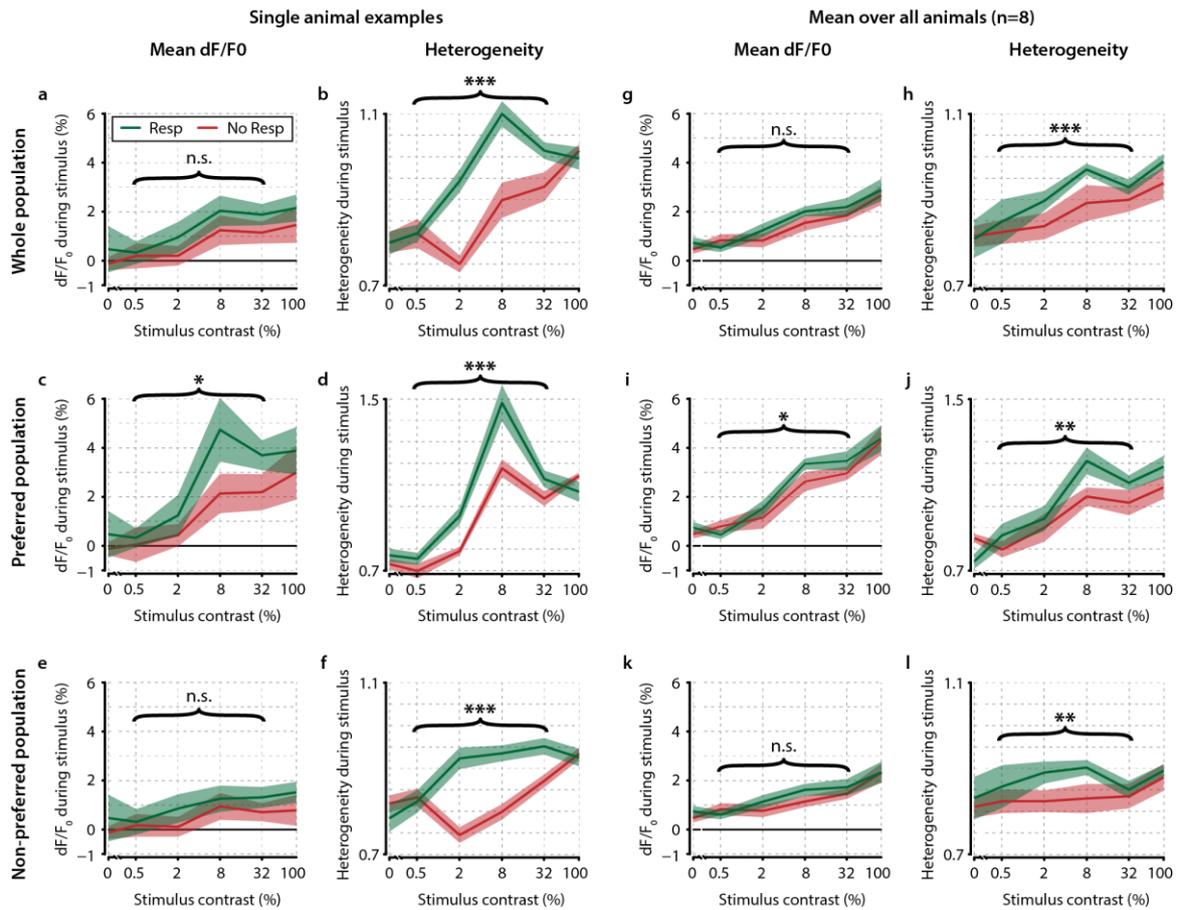

Figure supplement 3-1. Contrast-dependent responses of the preferred, non-preferred and whole population show distributed hit-correlated modulations in heterogeneity, but modulations in mean dF/F0 are smaller and only significant for the preferred population. **a-f**, Single animal neural response examples of the whole (**a,b**), preferred (**c,d**) and non-preferred (**e,f**) population of neurons. Mean dF/F0 shows hit-miss differences only within the preferred population ($p<0.05$) (**c**), but not within the population as a whole (**a**), nor within the non-preferred population (**e**) (paired t-tests, n.s.). However, heterogeneity shows significant differences across the whole population (**b**), as well as within the preferred (**d**) and non-preferred (**f**) population (paired t-test, $p<0.001$ for all comparisons). **g-l**, As **a-f**, but for across-animal comparison of hit-miss differences. All panels: Shaded areas show the standard error of the mean. Asterisks indicate statistical significance; * $p<0.05$; ** $p<0.01$; *** $p<0.001$.

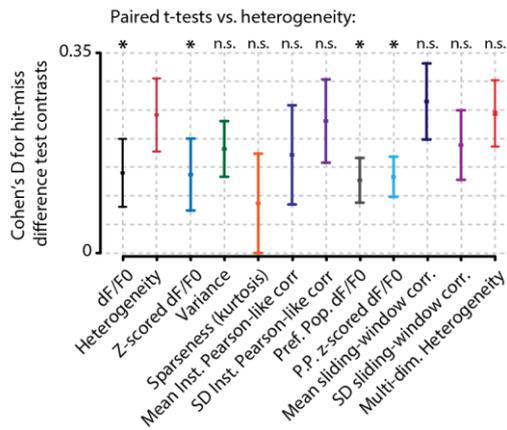

Figure supplement 3-2. Analysis of hit/miss effect size (Cohen's D) shows that simple average perform worst at separating hit from miss responses. Heterogeneity performed significantly better than mean whole-population (z-scored) dF/F0 (black and blue, FDR-corrected paired t-tests, p<0.05) and mean preferred population (z-scored) dF/F0 (grey and light blue, FDR-corrected paired t-tests, p<0.05). Variance (green), sparseness (orange), mean and SD of instantaneous Pearson-like correlations (dark blue and dark purple) and mean and SD of sliding window correlations (navy blue and light purple) did not differ significantly from heterogeneity. Error bars show the standard error of the mean. Asterisks indicate statistical significance; * p<0.05.

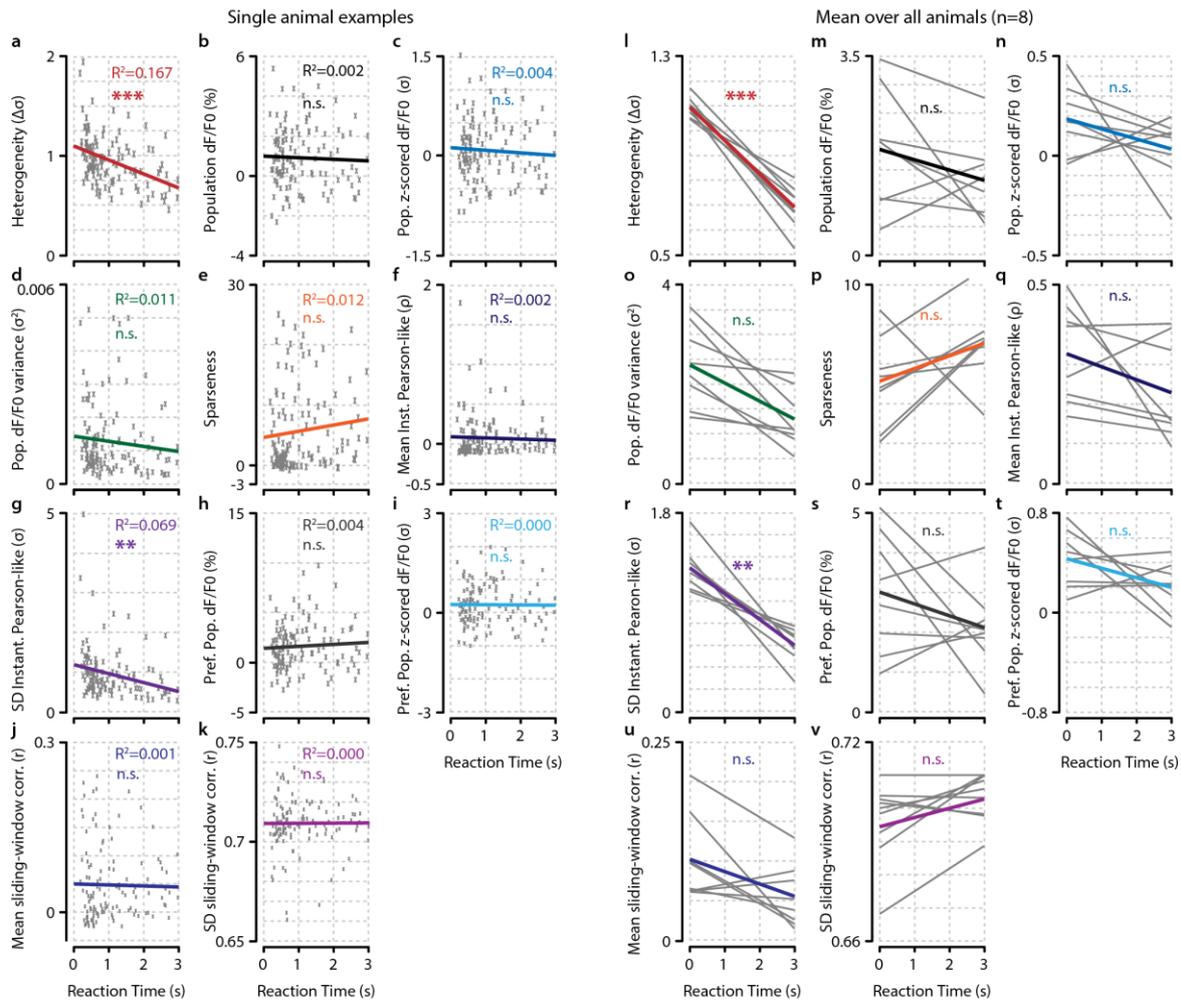

Figure supplement 4-1. Of several tested neural metrics, only heterogeneity and spread (SD) in instantaneous Pearson correlations show a significant relationship with behavioral reaction time (RT). **a-k**, RT-relationship of several metrics for example single animal. Each grey point represents a single trial. **a**, Heterogeneity plotted as a function of RT (Bonferroni-Holmes corrected regression slope vs. 0, p<0.001). **b**, population dF/F0 (n.s.). **c**, z-scored dF/F0 (n.s.). **d**, variance (n.s.). **e**, population sparseness (kurtosis) (n.s.). **f**, mean instantaneous Pearson correlation (n.s.). **g**, spread (SD) in instantaneous Pearson correlation (p<0.01). **h**, mean dF/F0 of preferred population (pref. pop.) (n.s.). **i**, mean z-scored dF/F0 of preferred population (P.P.) (n.s.). **j**, mean of sliding window correlations (n.s.). **k**, spread (SD) of sliding window correlations (n.s.). **l-v**, As **a-k**, but for analysis across animals. Each grey line represents the linear regression of a single animal. Non-grey lines represent the mean linear regression over all animals. Statistical significance of behavioral reaction time relationship was determined using Bonferroni-Holmes corrected one-sample t-tests of the slopes of all animals vs. 0. Heterogeneity ($p<0.001$) and spread in instantaneous Pearson correlation ($p<0.01$) were significant; all other metrics were not ($p>0.3$). All panels: Shaded areas show the standard error of the mean. Asterisks indicate statistical significance; * $p<0.05$; ** $p<0.01$; *** $p<0.001$.

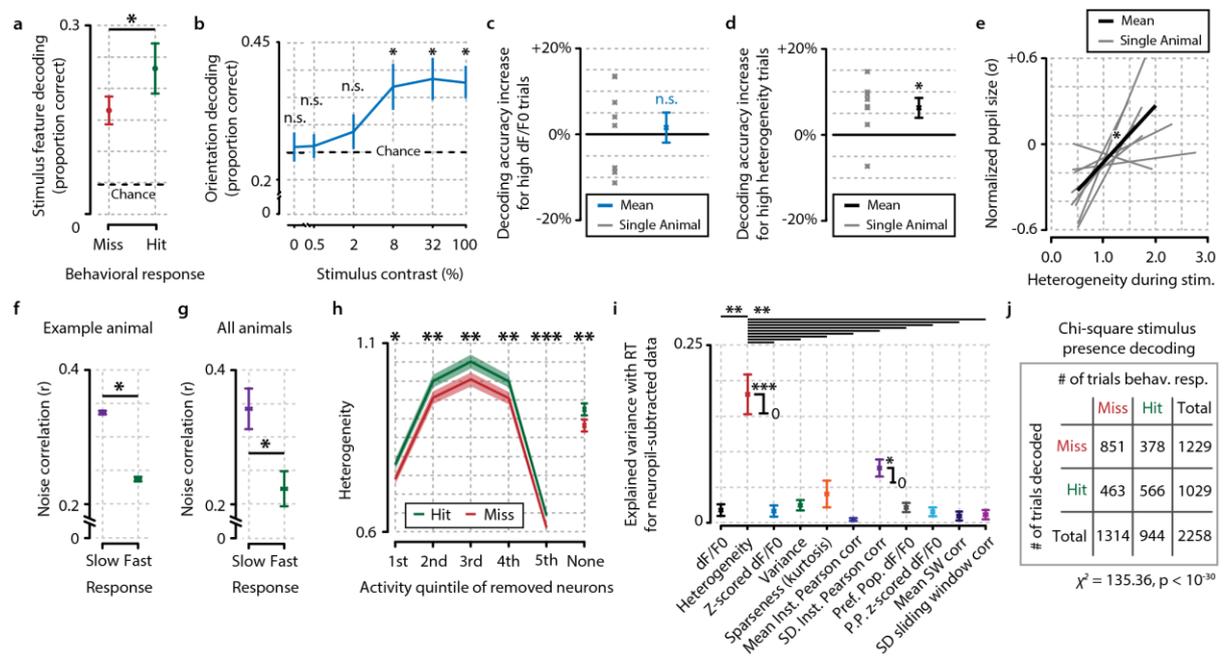

Figure supplement 4-2. Fidelity of stimulus feature representation and population heterogeneity are correlated with accurate visual detection and are not influenced by neuropil contamination. **a**, Analysis with a maximum-likelihood naive Bayes decoder shows that during hit trials stimulus features (orientation and contrast) are more accurately represented in the population response pattern than during miss trials (paired t-test, n=8 animals, p<0.05). **b**, Orientation decoding as a function of stimulus contrast shows a sigmoid curve. Statistical analysis revealed significantly above-chance orientation decoding for contrasts higher than 2% (post hoc FDR-corrected one-sample t-tests vs chance level (25%; four orientations were used); 0%, p=0.724; 0.5%, p=0.721; 2%, p=0.410; 8%, p<0.05; 32%, p<0.05, 100%, p<0.05). **c,d**, Orientation decoding accuracy does not increase when only strong responses to stimuli are taken into account (**c**, one-sample t-test, p=0.669), but does increase for high population heterogeneity (**d**, p<0.05). **e**, Mean pupil size 1s preceding stimulus onset is correlated with neuronal population heterogeneity during stimulus presentation, suggesting pre-stimulus arousal is related to heterogeneity (regression analysis per animal, one-sample t-test of slopes vs. 0, p<0.05, n=8 animals). **f**, Comparison of noise correlations (NCs) during slow and fast behavioral response trials for an example animal shows a significant reduction in NCs when the animal responds

fast (two-sample t-test, p<0.05, n=2211 pairs). **g**, Analysis across animals shows that this reduction is consistent (p<0.05, n=8 animals). **h**, Difference in heterogeneity between hit and miss trials is a population-distributed process and does not critically depend on selecting the most, or least, active neurons. On a single-trial basis we removed a single quintile of neurons within a z-scored activity bracket and recalculated the hit/miss difference after removal of this quintile (see Materials and Methods). While removal of the least (1$^{st}$ quintile) or most (5$^{th}$ quintile) active neurons per trial led to a decrease in absolute heterogeneity, the differences between hit and miss trials remained intact (paired t-tests hit vs. miss, n=8, 1$^{st}$ quintile p<0.05, 2$^{nd}$ p<0.005, 3$^{rd}$ <0.005, 4$^{th}$ p<0.005, 5$^{th}$ p<0.001, no removal p<0.005). Data show mean and standard error over animals (n=8). **i**, As fig. 4c, but when computed for neuropil-subtracted data (see Materials and Methods). Explained variance values were quite similar for most measures and the difference between heterogeneity and the other metrics appeared larger than without neuropil subtraction. All panels: Error bars and shaded areas indicate the standard error of the mean. Asterisks indicate statistical significance; * p<0.05; ** p<0.005; *** p<0.001. **j**, Chi-square analysis of stimulus presence decoding vs. behavioral response show a strong correspondence at single-trial level between the decoder's output and the animal's response (p<10$^{-30}$).

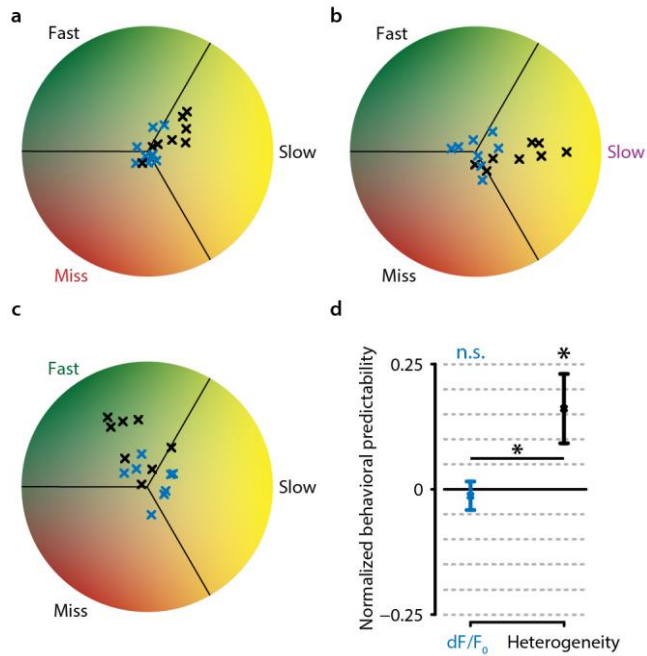

Figure supplement 5-1. Behavioral response is predictable on single trials when using heterogeneity, but not when using dF/F0. **a**, Predictive decoding of miss trials using mean population dF/F0 (blue) or heterogeneity (black) during 3s preceding stimulus onset. Each point is the mean prediction for a single animal (see Materials and Methods). **b**, same as **a**, but for slow trials. **c**, same as **a**, but for fast trials. **d**, Quantification of predictability shows chance-level prediction using dF/F0 (one-sample t-test, p=0.643), but above-chance prediction of behavioral responses using heterogeneity (p<0.05), as well as a significant difference in prediction performance between dF/F0 and heterogeneity (paired t-test, p<0.05). Points with error bars are mean +/- standard error (n=24; three points per animal (miss/slow/fast)).